\documentclass[a4paper,fleqn]{cas-sc}

\usepackage[numbers]{natbib}
\usepackage[english]{babel}
\usepackage{caption}
\usepackage{placeins}

\def\tsc#1{\csdef{#1}{\textsc{\lowercase{#1}}\xspace}}
\tsc{WGM}
\tsc{QE}
\tsc{EP}
\tsc{PMS}
\tsc{BEC}
\tsc{DE}

\DeclareMathOperator*{\argmin}{arg\,min}

\begin{document}
\let\WriteBookmarks\relax
\def\floatpagepagefraction{1}
\def\textpagefraction{.001}

\shorttitle{Hybrid Building Thermal Model for Building Energy Systems}

\shortauthors{Leandro Von Krannichfeldt et~al.}

\title [mode = title]{Combining Physics-based and Data-driven Modeling for Building Energy Systems}

%
\author[1,2]{Leandro Von Krannichfeldt}[type=editor,
                        orcid=0000-0001-8563-8086]

\cormark[1]


\ead{leandro.vonkrannichfeldt@epfl.ch}


\credit{Conceptualization, Methodology, Software, Investigation, Data curation,  Writing - original draft, Writing - review \& editing, Visualization}

\affiliation[1]{organization={EPFL},
    addressline={Route Cantonale}, 
    city={Lausanne},
    postcode={1015}, 
    country={Switzerland}}

\author[2, 3]{Kristina Orehounig}[%
   ]

\credit{Supervision, Conceptualization,  Writing - original draft, Writing - review \& editing, Resources}

\affiliation[2]{organization={Swiss Federal Laboratories for Materials Science and Technology (Empa)},
    addressline={Überlandstrasse 126}, 
    city={Dübendorf},
    postcode={8600}, 
    country={Switzerland}}

\affiliation[3]{organization={Vienna University of Technology (TUW)},
    addressline={Karlsplatz 13}, 
    city={Vienna},
    postcode={1040}, 
    country={Austria}}

\author%
[1]
{Olga Fink}
\ead{olga.fink@epfl.ch}

\credit{Supervision, Conceptualization, Writing - original draft, Writing - review \& editing, Resources}

\cortext[cor1]{Corresponding author}

\begin{abstract}
Building energy modeling plays a vital role in optimizing the operation of building energy systems by providing accurate predictions of the building's real-world conditions. In this context, various techniques have been explored, ranging from traditional physics-based models to data-driven models. Recently, researchers are combining physics-based and data-driven models into hybrid approaches. This includes using the physics-based model output as additional data-driven input, learning the residual between physics-based model and real data, learning a surrogate of the physics-based model, or fine-tuning a surrogate model with real data. However, a comprehensive comparison of the inherent advantages of these hybrid approaches is still missing. The primary  objective of this work is to evaluate four predominant hybrid approaches in building energy modeling through  a real-world case study, with focus on indoor thermodynamics. To achieve this, we devise three scenarios reflecting  common levels of building documentation and sensor availability, assess  their performance, and analyze their explainability using hierarchical Shapley values. 
The real-world study reveals three notable findings. 
First, greater  building documentation and sensor availability lead to higher prediction accuracy for hybrid approaches. 
Second, the performance of hybrid approaches depends on the type of building room, but the residual approach using a Feedforward Neural Network as data-driven sub-model performs  best on average across all rooms. This hybrid approach also demonstrates a superior ability to leverage the simulation from the physics-based sub-model. Third, hierarchical Shapley values prove to be an effective tool for explaining and improving hybrid models while accounting for input correlations.

\end{abstract}


\maketitle

\section{Introduction}
\begin{figure}[h]
	\centering
	\includegraphics[width=1.0\linewidth]{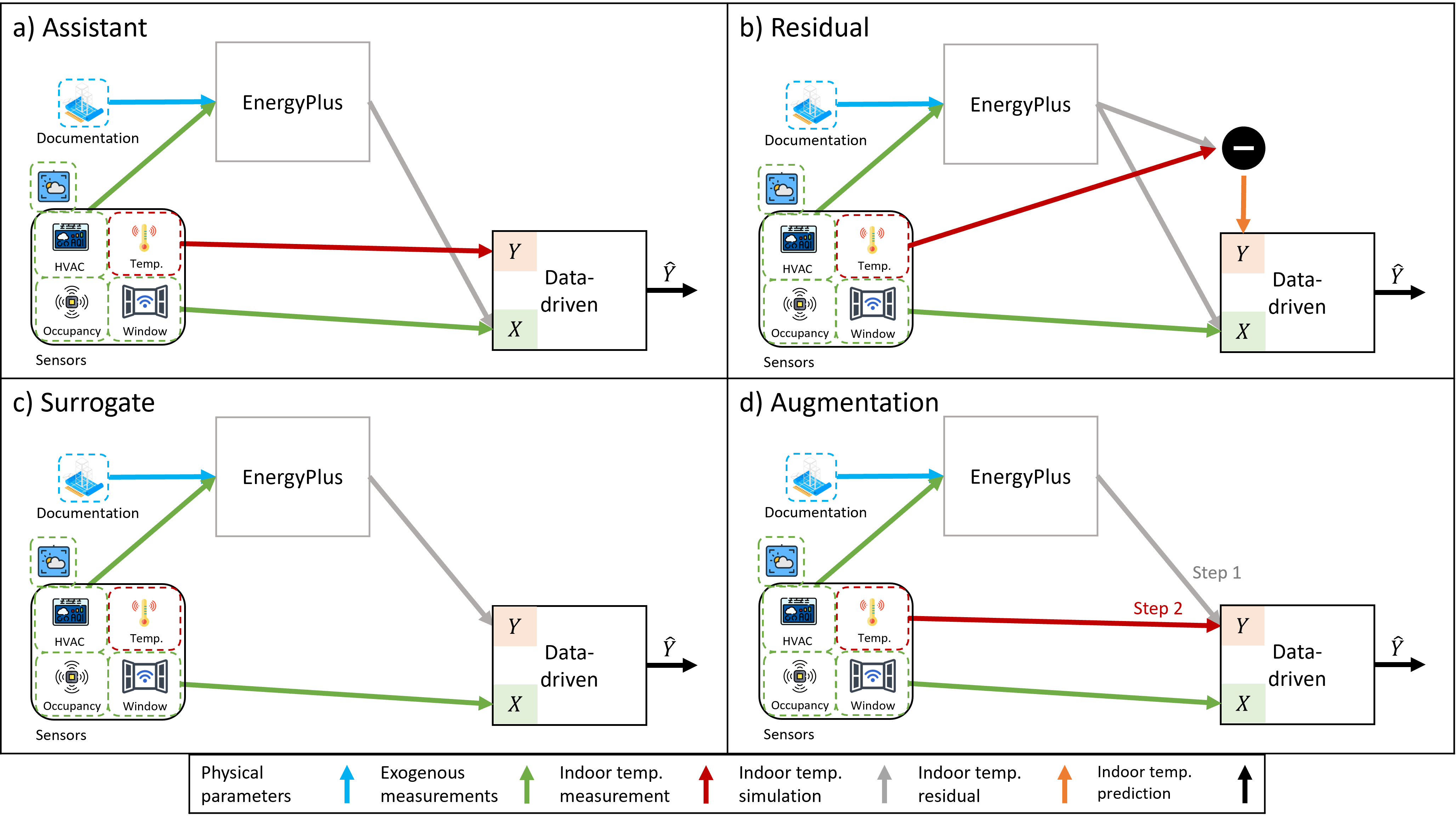}
	\caption{Overview of the different hybrid approaches with physics-based EnergyPlus and data-driven combination. The data sources are indicated as building documentation and sensors with various sensor groups. The corresponding data flows are indicated by colored arrows, which meanings are given in the legend at the figure bottom. The green exogenous data arrow indicates all the sensor measurements apart from the indoor room temperature. $X$ and $Y$ denote the input features and target variables of the data-driven model for learning purposes. $\hat{Y}$ represents the hybrid model's indoor temperature prediction. Note that the augmentation approach involves a two-step learning procedure. While step 1 indicates the learning on simulated data, step 2 denotes the fine-tuning on real indoor temperature.}
	\label{fig:hybrid}
\end{figure}

In the last decade, the operation of buildings accounted for approximately 30\% of global energy consumption and 26\% of CO\textsubscript{2} emissions \cite{iea_buildings}. Over half of this consumption stems from the operation of building energy systems such as Heating, Ventilation and Air Conditioning (HVAC), as well as the electrical systems. This situation underscores the urgent need for measures to reduce energy consumption in the operation these systems. Fortunately, advancements in digitalization and sensor deployment provide a foundation for using analytical and machine learning tools to optimize the performance  of building energy systems. On this basis, Building Energy Models (BEM) can be developed  to simulate a building's real-world condition and predict  future behaviour thereby enabling the recommendation of optimal control actions. However, a major challenge remains:  the availability of building information and sensor data, particularly in terms of comprehensive building documentation, sensor coverage, and the length of recorded data. This lack of building information and sensor data complicates the reliable and accurate construction of BEMs but also hinders the widespread adoption of energy optimization strategies in buildings. \\
The Building Energy Model is a digital representation of a building designed  for simulating energy and thermodynamics. It is constructed  based on various  characteristics of the building, including building geometry, material properties, installed energy systems, and operational  inputs such as weather conditions, HVAC operation, and occupancy schedules \cite{deb_review_2021}. To accurately reflect real-world conditions, the BEM is initially calibrated with sensor data to ensure faithful representation of the building and its energy systems. Once  calibrated, the BEM can predict building operational performance and indoor temperature evolution across different time scales such as quarter-hourly or hourly resolution. Its comprehensive predictive capabilities make the BEM a versatile tool for various applications in building planning and operations. In retrofitting studies, it serves as an evaluation tool to assess potential measures for improving  energy efficiency \cite{shen_feasibility_2019}. In building operations, it provides detailed temperature predictions as basis for control algorithms, enabling the optimization of energy system operations \cite{khayatian_benchmarking_2023}. \\
Classical BEMs are physics-based models that use a system of differential equations to represent all building subsystems and their interactions. This includes equations for indoor heat balance, HVAC dynamics, and incident solar radiation. These models, often referred to as physics-based models, provide insights into the underlying physical phenomena governing building performance \cite{deb_review_2021}. However, the accuracy of these simulations can be compromised by the unavailability of detailed building information. Additionally, setting up a BEM can be a time-consuming process that requires significant  expertise. A widely used physics-based modeling software in both academia and industry is EnergyPlus \cite{energyplus}, which enables  dynamic thermal simulations and supports  the energy-efficient design and operation of buildings. In case of simplified building descriptions or the requirement to lower computational complexity, reduced-order physics-based models may be employed. These models consist of simplified physical descriptions in the form of differential equations paired with data-driven identification of model coefficients, such as thermal Resistance-Capacitance (RC) models \cite{deb_review_2021}.\\
More recent approaches of building energy modeling focus on data-driven methods, which utilize  statistical and machine learning techniques such as Autoregressive Integrated Moving Average (ARIMA), Feedforward Neural Network (FFNN), Long Short-Term Memory (LSTM) and Convolutional Neural Network (CNN) \cite{deb_review_2021}. These methods rely on sensor measurements to establish relationships between defined input and output variables. Commonly modelled relationships include those between building operation data -- such as energy consumption or temperature -- and the corresponding sensor readings. Since these models are applied to directly data, their underlying mechanisms are not easily accessible, classifying them as data-driven models. Furthermore, their complexity reduces interpretability. In such cases, post-hoc explainability methods like Shapley values \cite{SHAP}, can be used to analyze model predictions. Unlike physics-based models, data-driven approaches do not require detailed physical building information or the  calibration of physical parameters, making them easier to set up as digital twins. However, their performance is highly dependent on the quantity and quality of the available measurement data. Despite this dependency, data-driven models have demonstrated high accuracy in numerous studies and are particularly well-suited for modeling buildings at an urban scale, thanks  to their reduced configuration time \cite{pan_building_2023}. \\
An emerging category  of approaches for building energy modeling is known as hybrid or physics-induced modeling, which combines elements from physics-based and data-driven methods. These approaches offer significant advantages, particularly in scenarios where sensor data or detailed documentation is lacking.\\
Several research studies aim to combine reduced-order physics-based models with data-driven models through  two general strategies. 
The first strategy involves formulating a loss function based on the reduced-order model to train the data-driven model. A prominent example is the use of Physics-informed Neural Network (PINN), where a physics-informed loss function is devised based on RC model thermodynamics in order to train a FFNN \cite{gokhale_physics_2022, nagarathinam_pacman_2022, chen_physics-informed_2023}. 
The second strategy consists of incorporating reduced-order model elements into a data-driven model architecture. A number of works model building dynamics with a State-Space model framework and parameterize state and/or observation equation with FFNN \cite{drgona_physics-constrained_2021}, Neural Ordinary Differential Equations \cite{taboga_neural_2024} or Graph Neural Networks \cite{yang_physics-constrained_2024}. However, ensuring physical consistency still remains a challenge in aforementioned strategies \cite{di_natale_towards_2023}. \\
Another line of work researches  the combination of high-fidelity physics-based models with  data-driven models. We identify four distinct approaches in the context of our research: assistant, residual, surrogate, and augmentation. In the \textbf{assistant strategy}, the output from a physics-based model is used as an additional input to the data-driven model \cite{chen_hybrid-model_2022, alden_digital_2022} or conversely, the data-driven model can provide inputs or corrections to the physics-based model \cite{verma_ann_2021, causone_data-driven_2019, lin_hybrid_2021}.
This additional input may provide valuable context information, but also increases the number of input features. The \textbf{residual strategy} involves using a data-driven model to learn the residuals between the output of  a physics-based model and actual observed data \cite{sterling_improving_2014, cui_hybrid_2019, dong_hybrid_2016, massa_gray_hybrid_2018}. This approach aims to capture unmodelled physical phenomena  and variations in the data that the physics-based model may not fully account for. It is particularly useful when certain  inputs are not reliably represented in the data or when  incorporating domain knowledge with a physical reference input is necessary. In the \textbf{surrogate strategy}, a data-driven model is trained to replace the physics-based model by using the same inputs, with the physics-based model's simulations serving  as the target outputs \cite{westermann_using_2020, westermann_using_2021, edwards_constructing_2017, veiga_application_2021, sharif_developing_2019}. The primary motivation for this approach is to reduce computation time, allowing the data-driven model to perform simulations that would otherwise be time-consuming with the physics-based model. Nevertheless, it relies on a sufficiently accurate simulator. The \textbf{augmentation strategy} augments the real data with simulated data from a physics-based model, training the data-driven model on this augmented dataset and subsequently adapting it to real-world situations \cite{nutkiewicz_data-driven_2018, choi_context-aware_2020, ahn_prediction_2022, vaghefi_hybrid_2016}. This augmentation strategy is particularly advantageous when  little to no real-world data is available, as it allows the model to leverage simulated data to improve its accuracy and generalization capabilities. However, this approach may suffer from performance degradation  if there is a significant disparity between simulated and real data. An overview of the different characteristics of hybrid approaches in the context of our research is shown in Figure \ref{fig:hybrid}.\\
Several studies  investigate the dependency  of hybrid models on documentation and sensor measurements. For example, various levels of building documentation are explored in \cite{chen_hybrid-model_2022} using  an assistant methodology that combines IDA-ICE software and Gradient Boosting Regression Trees (GBRT). In scenarios with limited sensor data, \cite{martellotta_use_2017} employs  a Surrogate-FFNN model based on EnergyPlus  simulations to assess performance with varying data availability -- 100\%, 30\% and 20\%, or even less \cite{chen_utilizing_2024}. Additionally, \cite{jain_transfer_2021} explores the impact of seasonally limited data using an Augmentation-FFNN-CNN model implemented through  the MATLAB toolbox CARNOT \cite{carnot}.
Few research studies address  the explainability of hybrid models. For example, \cite{yang_enhancing_2021} evaluates a Surrogate-Random Forest model trained on simulated data from EnergyPlus using Pearson correlation and  Gini importance scores  to assess feature relevance. Similarly, \cite{chen_introducing_2022} enhances the interpretability of a Surrogate-GBRT model trained on simulated data  from EnergyPlus by incorporating a causal inference framework, thereby making the hybrid model inherently more transparent. \\
Although hybrid approaches in BEM are gaining increasing attention, three significant  research gaps remain.
First, most  recent studies focus  on developing new hybrid methods within  specific  data contexts,  yet there is a lack of  a comprehensive research  comparing the advantages and disadvantages of different hybrid approaches. A systematic evaluation of these methods is still missing, which limits our understanding of their relative strengths and weaknesses. Second, there are only few works that compare scenarios with limited building documentation or sensor data. 
Third, the majority of existing literature primarily evaluates hybrid models in terms of accuracy, with less attention given to the explainability of these approaches. Greater emphasis on explainability is crucial for understanding the hybrid models general behaviour, uncovering model biases of the physics-based sub-model and building trust for real-world application. Moreover, due to environmental and operational physical dependencies, sensor measurements exhibits interaction effects that are reflected  in the model features. As a consequence, the joint influence of feature groups and non-linear interactions  between features impact  model predictions.\\

These research gaps highlight the need for more comparative studies and a deeper exploration of how hybrid BEMs perform under varying levels of data availability and quantity. We address these research gaps by conducting  a comprehensive study of hybrid approaches in a real-world setting. For explainability, we opt for post-hoc explanation of model predictions, given the complexity of the investigated models. We introduce  the hierarchical Shapley value as a suitable framework for our setting, as it effectively assigns feature contributions to correlated feature groups within the original output unit.
In our study, we focus on the thermal modeling aspect of the BEM to predict indoor thermodynamics for two reasons. Firstly, the understanding of thermodynamics forms the foundation for all energy related calculations, also in physics-based models. Secondly, the focus on thermodynamics allows for a clear effect modeling and facilitated model analysis. Furthermore, we concentrate on a high-fidelity physics-based model for the construction of hybrid models.
This paper tries to address the previously mentioned research gaps by making the following three main contributions:

\begin{itemize}
  \item We enhance the understanding of  hybrid building energy models by investigating and comparing four predominant hybrid approaches across  three challenging real-world scenarios, each characterized by varying levels of  building documentation and sensor data availability.
  \item We apply a hierarchical Shapley value framework to an agglomerative clustering analysis using Pearson's distance metric, providing valuable  insights into the nature of hybrid models while accounting for the correlations. This also allows to investigate potential model biases of the physics-based part such as a bias at higher outdoor temperatures.
  \item We examine  and compare performance of the four hybrid approaches in a limited training data setting, offering a detailed  analysis of their dependency on data quantity and their robustness under constrained conditions.
\end{itemize}

The remainder of this paper is organized as follows: Section 2 
outlines  the methodology used in this study. Section 3 introduces
the data set-up and implementation details of the hybrid approaches. Section 4 conducts case studies across  various  documentation and sensor scenarios and analyzes the results. Finally, Section 5 draws conclusions and provides  an
outlook on future research directions.

\section{Methodology}
Our methodology consists of combining a physics-based EnergyPlus model with a data-driven model in four prevalent combinations. We then proceed to evaluate these combinations based on accuracy using  standard accuracy metrics, explainability using hierarchical Shapley values, and data dependency under different documentation and sensor availability scenarios.

\subsection{Hybrid model}

In our research, we evaluate the four most prevalent forms of combining physics-based and data-driven models in building energy modeling, shown in Figure~\ref{fig:hybrid}. The \textbf{assistant approach} incorporates the  indoor temperature predictions from the physics-based model as an additional input to the data-driven model. The \textbf{residual approach} involves using a data-driven model to learn the indoor temperature residuals -- the differences between the physics-based model's predictions and the real indoor temperatures. The \textbf{surrogate approach} aims  to train a data-driven model to fully replace the physics-based model for indoor temperature prediction, taking the physics-based simulation as training label. Finally, the \textbf{augmentation approach} takes this step further by fine-tuning the data-driven model with real data following its initial pre-training as  a surrogate.

The corresponding algebraic equations for these hybrid models are given as:

\begin{align}
    \mathbf{f}_{assistant}(\mathbf{x}) &= \mathbf{f}_{dd}(\mathbf{x}, \mathbf{f}_{EP}(\mathbf{x})), \quad l_{dd}(\mathbf{y}, \mathbf{f}_{assistant}(\mathbf{x}))\\
    \mathbf{f}_{residual}(\mathbf{x}) &= \mathbf{f}_{dd}(\mathbf{x}, \mathbf{f}_{EP}(\mathbf{x})), \quad l_{dd}(\mathbf{y} - \mathbf{f}_{EP}(\mathbf{x}), \mathbf{f}_{residual}(\mathbf{x})) \\
    \mathbf{f}_{surrogate}(\mathbf{x}) &= \mathbf{f}_{dd}(\mathbf{x}, \mathbf{f}_{EP}(\mathbf{x})), \quad l_{dd}(\mathbf{f}_{EP}(\mathbf{x}), \mathbf{f}_{surrogate}(\mathbf{x})) \\
    \mathbf{f}_{augmentation}(\mathbf{x}) &= \mathbf{f}_{surrogate}(\mathbf{x}, \mathbf{f}_{EP}(\mathbf{x})), \quad l_{dd}(\mathbf{y}, \mathbf{f}_{augmentation}(\mathbf{x}))
\end{align}

where $\mathbf{f}_{EP}$ and $\mathbf{f}_{dd}$ are the physics-based EnergyPlus and data-driven prediction functions, $\mathbf{x}$ is a sample with the corresponding feature dimension, $\mathbf{y}$ is the target with real indoor temperatures, and $l_{dd}$ is the data-driven loss function. We employ all four approaches in our case study and compare them extensively.

\subsection{Physics-based models}
\begin{figure}[h]
	\centering
	\includegraphics[width=0.5\linewidth]{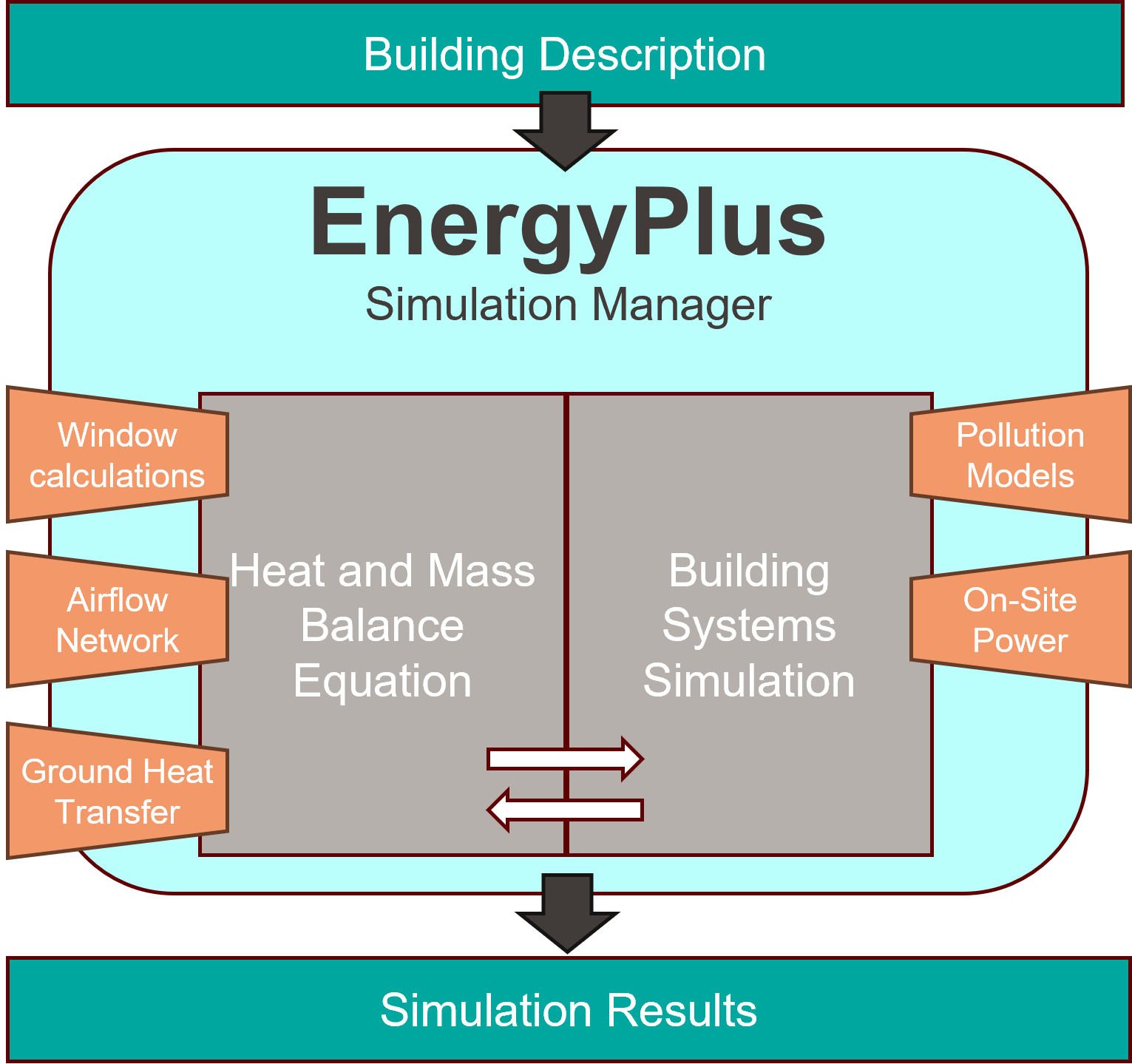}
	\caption{Overview of the EnergyPlus workflow with modules, adapted from \cite{energyplus}.}
	\label{fig:energyplus}
\end{figure}

As our physics-based model, we utilize the widely recognized open-access building simulation software EnergyPlus (EP) \cite{energyplus} to construct a comprehensive building energy model, as depicted in Figure~\ref{fig:energyplus}. EnergyPlus simulates the building and its subsystems using a system of differential equations. These modules include equations for the building envelope, indoor heat balance, mass balance, HVAC fluid dynamics, lighting system, window performance, and weather conditions. To set-up the EP model, the building geometry and physical parameters are curated from the building documentation and provided in an Intermediate Data Format file. Alongside  weather data formatted in the EnergyPlus weather file, these inputs allow the internal simulation manager to accurately simulate the building's dynamics behaviour. Once the multi-zone building model is set up, its parameters are calibrated by aligning the real target variable with the simulation output using the sensor data from the building and individual rooms. This procedure ensures accurate simulation of thermodynamics within the building. In case of less detailed building information, an archetypal EnergyPlus model can be constructed. For this purpose, we use the archetype framework Cesar-P \cite{cesar, cesarp} for automated model construction. On the basis of building footprint, height, year of construction and building type, the Cesar-P framework constructs a suitable archetypal Energyplus model with pre-defined material and geometry settings.

\subsection{Data-driven models}
In the case of data-driven models we investigate Linear Regression (LR), Feedfoward Neural Network (FFNN) and Random Forest (RF).
We choose the LR due to its simplicity as well as strong baseline performance. The FFNN is chosen for its popularity and adequacy for regression problems. It also has computational advantages compared to kernel-based methods such as Support Vector Regression. The RF is selected as a representative of the tree ensemble models that are widely used in regression problems for their accuracy. Moreover, both FFNN and RF are able to represent the building thermodynamics of all rooms within a single model. A Recurrent Network architecture is not considered since no time-lagged features are used as input. This enables a direct comparison to the pure physics-based EnergyPlus simulation, which doesn't allow explicit time lag integration.\\
The multiple \textbf{Linear Regression} is defined by the Least Squares optimization problem \cite{bishop_pattern_2006}:
\begin{equation}
    \min_\mathbf{W} || \mathbf{W} \mathbf{X} - \mathbf{Y} ||^2
\end{equation}
The solution  to this problem is given by:
\begin{equation}
    \mathbf{W} = (\mathbf{X}^T\mathbf{X})^{-1}\mathbf{X}\mathbf{Y}
\end{equation}

where $\mathbf{X} \in \mathbb{R}^{N \times d}$ is the feature matrix consisting of $N$ samples and $d$ features, $\mathbf{Y} \in \mathbb{R}^{N \times K}$ is the target matrix composed by $N$ samples and $K$ target variables, and $\mathbf{W} \in \mathbb{R}^{d \times K}$ are the regression coefficients. It is important to note that the solution to the Multiple Linear Regression decouples between the target variables into individual Linear Regression solutions per target. Thus, the prediction function for a target room temperature $\mathbf{y}^{(k)}$ given a sample $\mathbf{x}$ is:
\begin{equation}
    \mathbf{f}(\mathbf{x}) = \mathbf{W}_k\mathbf{x}
\end{equation}
Regarding the \textbf{Feedforward Neural Network}, the model equations are defined by:
\begin{align}
    \mathbf{f}(\mathbf{x}) &= \mathbf{z}^{(L)} \circ \mathbf{z}^{(L-1)} \circ ... \circ \mathbf{z}^{(1)}(\mathbf{x}) \\
    \mathbf{z}^{(k)} &= \alpha^{(k)}(\mathbf{W}^{(k)}\mathbf{z}^{(k-1)} + \mathbf{b}^{(k)}) \\
    \mathbf{z}^{(1)} &= \alpha^{(1)}(\mathbf{W}^{(1)}\mathbf{x}+ \mathbf{b}^{(1)})
\end{align}

where, $\alpha^{(i)}$, $\mathbf{W}^{(i)}$, and $\mathbf{b}^{(i)}$ represent  the activation function, network weights, and bias of the $i$-th layer, respectively. Further, the $\circ$-symbol denotes the composition of functions. The network weights are learned  by using an optimizer to iteratively solve the optimization problem with stochastic approximation \cite{bishop_pattern_2006}:

\begin{equation}
    \mathbf{W} = \argmin_{\mathbf{W}} \dfrac{1}{N} \sum_{i=1}^{N} || \mathbf{y}_i - \mathbf{f}(\mathbf{x}_i, \mathbf{W})||^2
\end{equation}

The \textbf{Random Forest} regression works by averaging many unbiased tree models to reduce the overall model variance \cite{bishop_pattern_2006}. By denoting a regression tree as $T_b$, the model equation for $B$ number of trees is given as:

\begin{equation}
    \mathbf{f}(\mathbf{x}) = \dfrac{1}{B} \sum_{b=1}^B T_b(\mathbf{x}), \quad \quad T_b(\mathbf{x}) = \sum_{j=1}^J \mathbf{c}_j \mathbb{1}_{\{\mathbf{x} \in R_j\}}
\end{equation}

where $R_j$ is a region in the feature space partition, $\mathbf{c}_j$ are the constant values for the region $R_j$ and $\mathbb{1}_{\{\cdot\}}$ represents the indicator function. The parameters are found by minimizing the following combinatorial optimization problem:
\begin{equation}
    \mathbf{R}, \mathbf{c} = \argmin_{\mathbf{R}, \mathbf{c}} \sum_{j=1}^J \sum_{\mathbf{x}_i \in R_j} ||\mathbf{y}_i - \mathbf{c}_j||^2
\end{equation}

\subsection{Explainability}
We refer to explainability as the ability to describe and justify a model’s predictions, especially for complex data-driven models, such as Deep Learning models used in this research \cite{marcinkevics_interpretable_2023}.
In a combination of physics-based and data-driven models, the explainability of the data-driven model as well as of the interdependencies between the two sub-models is crucial for the overall understanding. In particular, we want to examine these two aspects through the lens of feature contributions for model predictions on a global level. In this way, we can assign feature importance to get a better understanding of the model's general behaviour, uncover model biases and potentially give suggestions whether to record particular sensor variables. To comprehensively explain the hybrid models, we employ a Shapley value framework \cite{shapley_value_1953} with the Shapley additive explanations (SHAP) package \cite{SHAP}. Given the challenge of correlated input features, we opt for a hierarchical Shapley value calculation \cite{owen_values_1977}. In this approach, hierarchical Shapley values are recursively computed based on a predefined  hierarchy. We establish this hierarchy using agglomerative clustering \cite{Agglomerative}, with Pearson's distance \cite{fulekar_bioinformatics_2009} as the distance metric. Consequently, the distance matrix $\mathbf{D}$ can be expressed  as follows:

\begin{equation}
	\mathbf{D} = 1 - \mathbf{R}_{\mathbf{X} \mathbf{X}}, \quad \mathbf{R}_{\mathbf{X} \mathbf{X}}^{(i,j)} = | cov(\mathbf{X}_i, \mathbf{X}_j) (\sigma_i \sigma_j)^{-1} |
\end{equation}

where each entry of the absolute correlation matrix $\mathbf{R}_{\mathbf{X} \mathbf{X}}$ represents the absolute value of the Pearson correlation between two feature columns. Here, $cov$ denotes covariance and $\sigma$ represents standard deviation.
Subsequently, the hierarchical Shapley value $\phi_i$ of a specific feature $i$ is computed  as follows:

\begin{equation}
    \phi_i(v, B)=\sum_{R \subseteq M \backslash\{k\}} \sum_{T \subseteq B_k \backslash\{i\}} \frac{1}{|M| |B_k|} \frac{1}{\binom{|M|-1}{|R|}} \frac{1}{\binom{|B_k|-1}{|T|}} [v(Q \cup T \cup\{i\})-v(Q \cup T)]
\end{equation}

where the norm  $|\cdot|$ in this context denotes the size of a set, $M$ is the set of clusters, $B = \{B_1, ..., B_m\}$ represents the partition of features into clusters, $B_k$ is the cluster containing feature $i$ and $k$ is the index of cluster containing feature $i$. $R$ is a subset of clusters and the first binomial coefficient calculates the corresponding number of ways that to choose $|R|$ subsets out of $|M| - 1$ clusters. $T$ is a subset of features and the second binomial coefficient calculates the corresponding number of ways that to choose $|T|$ features within cluster $B_k$ from $|B_k| - 1$ features. $Q = \bigcup_{r \in R} B_r$ is a union of features in the clusters defined by $R$, and $v(\cdot) = \mathbb{E}[\mathbf{f}(\mathbf{X}) | \mathbf{X}_{(\cdot)}]$ is the expected prediction of a subset. Therefore, the term $v(Q \cup T \cup\{i\})-v(Q \cup T)$ describes the marginal contribution of feature 
$i$ when added to the union $Q \bigcup T$.
In addition to the hierarchical Shapley values, we analyze the intrinsic interpretability of different models: Linear Regression through regression coefficients, Random Forest through Gini importance, and Neural Network through Jacobian matrix sensitivity. However, these intrinsic measures become less interpretable in the presence of feature interactions from correlated input features.

\subsection{Evaluation metrics}

For evaluating the accuracy of temperature predictions, we utilize the widely used metrics: Mean Absolute Error (MAE) for its insensitivity to outliers, Mean Absolute Percentage Error (MAPE) for its comparability across rooms as well as Root Mean Squared Error (RMSE) for its outlier penalization. These metrics are defined as follows:

\begin{equation}
	\text{MAE} = \dfrac{1}{T} \sum_{t=1}^{T} |y_t - \hat{y}_t|
\end{equation}

\begin{equation}
	\text{MAPE} = \dfrac{1}{T} \sum_{t=1}^{T} \left| \dfrac{y_t - \hat{y}_t}{y_t} \right|
\end{equation}

\begin{equation}
	\text{RMSE} = \sqrt{\dfrac{1}{T} \sum_{t=1}^{T}(y_t - \hat{y}_t)^2}
\end{equation}

where $y_t$ and $\hat{y}_t$ represent the real and predicted room temperatures at time $t$, across $T$ time steps.

\section{Case study}
\subsection{Dataset}

\begin{figure}[h]
	\centering
	\includegraphics[width=0.65\linewidth]{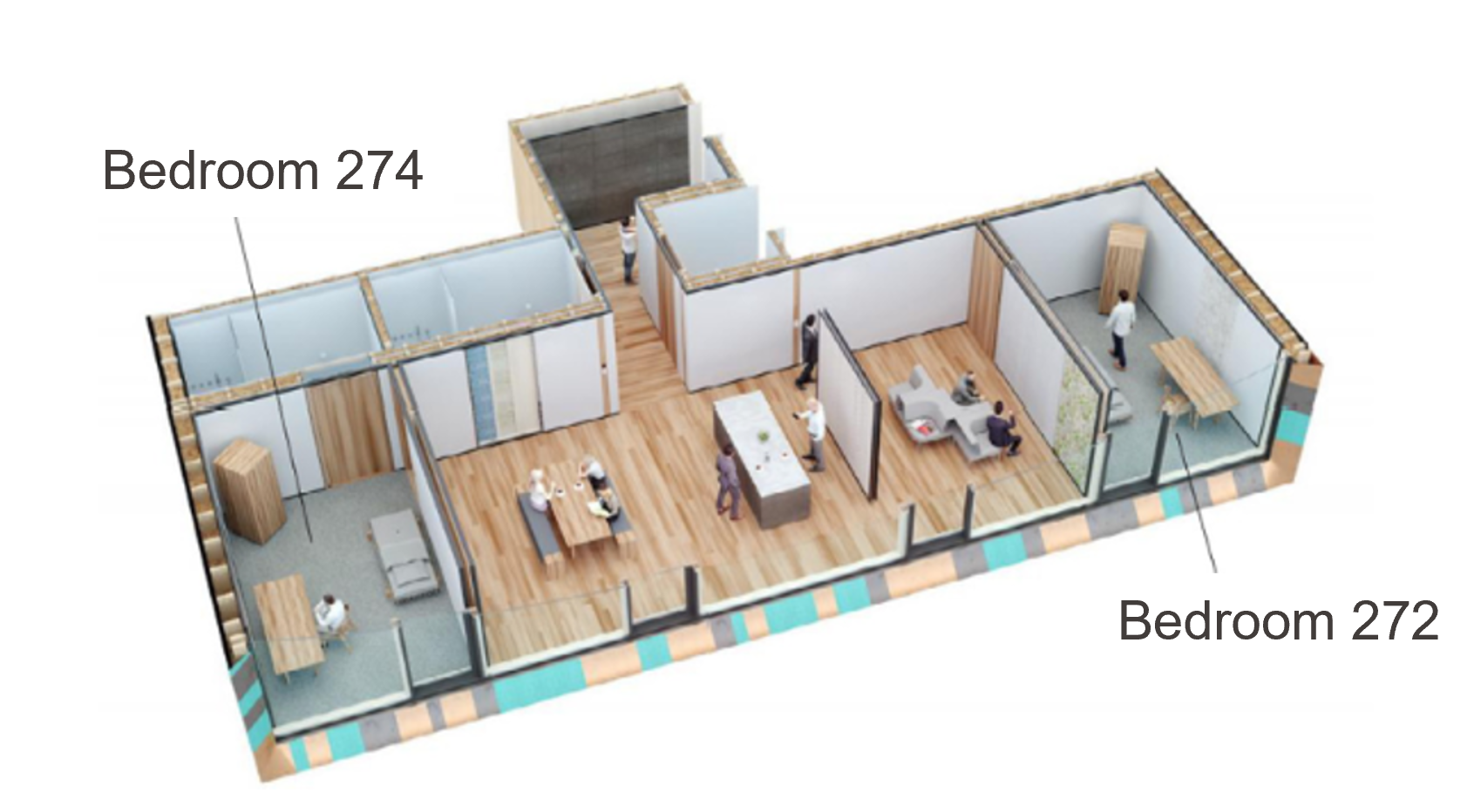}
	\caption{UMAR unit at Empa with two bedrooms, one living room and two bathrooms.}
	\label{fig:umar}
\end{figure}

The main subject of investigation is the inhabited experimental unit Urban Mining and Recycling (UMAR), located at the Swiss Federal Laboratories for Materials Science and Technology (Empa) in Dübendorf \cite{richner_nest_2018}, as shown in Figure~\ref{fig:umar}. 
In terms of representativeness, previous research indicates that the UMAR unit is representative of the average Swiss residential building stock around 2050 \cite{mutschler_benchmarking_2021}. The unit comprises several rooms equipped  with various indoor and outdoor sensors, which have been recording data over several years. Our study concentrates on the five most frequently used rooms: two bedrooms (R272, R274), a living room (R273), and two bathrooms (R275, R276). The utility room is not regarded for our study. A comprehensive  overview of all  weather, building, and room sensors utilized in this study is provided  in Table~\ref{table:features}. The dataset for our case study is recorder at a 1-minute resolution for all sensor variables spanning the two years 2020 and 2021. For the whole dataset, the percentage of missing values is below 1\%. After linearly interpolating the missing values, the dataset is aggregated to 15-minutes resolution. The dataset is split into a 50\%/50\% training/test set, with the year 2020 used for training and the year 2021 used for testing. 

\subsection{Documentation/sensor scenarios}
The case study aims to investigate the performance of hybrid models in common real-world scenarios. To achieve this, we define commonly encountered levels of documentation (floor plan, construction information, ...) and sensor availability, as outlined in Table~\ref{table:scenarios}.
The documentation level is categorized as either archetypal, using general information such as building layout, type, glazing ratio, and year of construction, or detailed, relying  on comprehensive  building documentation. Sensor availability is classified into three levels: weather, building and room, also detailed in Table~\ref{table:features}. The weather-level  includes only weather-related variables such as outdoor temperature and solar radiation. The building-level adds  measurements for entire  building, including total heating and cooling power. The room-level further includes room-specific  variables such as temperature setpoints, occupancy, and shading.
Each specific combination of documentation and sensor levels is referred to as a scenario. In our study, we focus on three typical scenarios: W-, WB- and WBR-scenario. 
In the W-scenario, we assume that only weather-related sensors are available as well as only basic building information. On this basis, we utilize the archetype framework Cesar-P \cite{cesarp} construct a simplified EnergyPlus model as physics-based model.
In the WB-scenario, we consider having weather and general building related sensors as well as detailed construction information. In this case, we use an uncalibrated detailed EnergyPlus model, since a precise calibration without room sensors is not possible.
In the WBR-scenario, we presume having access to weather, building and room sensors as well as detailed construction information. Here, we take a calibrated detailed EnergyPlus model to serve as the physics-based component in the hybrid models.

\subsection{Model architecture}
In our time series regression framework, we use all exogenous variables to predict the indoor temperatures of all rooms at the same time step. No time-lagged data is used as input to enable a direct comparison to the pure physics-based EnergyPlus simulation. Regarding the physics-based model, we use a Cesar-P model for the W-scenario and a detailed EnergyPlus model for the WB- and WBR-scenarios.
The Cesar-P model is configured using parameters such as  building footprint, type, age, heating system type, and glazing ratio and set up as a simplified EnergyPlus model. For the detailed EnergyPlus model, we employ the same model as described in \cite{khayatian_benchmarking_2023}, using an uncalibrated version for the WB-scenario and a fully calibrated version according to the authors's calibration cycle for the WBR-scenario. In both instances, each room is treated  as a single temperature zone.\\
For the data-driven part, we compare three models of differing complexity: LR, FFNN and RF.  While LR and RF implementations are used from Scikit-learn \cite{sklearn}, the FFNN is implemented in Pytorch \cite{pytorch}. All three models are trained on standardized feature inputs. The LR model used is a multiple ordinary least squares regression with an intercept. The FFNN consists of two layers with 128 neurons activated by the sigmoid function. It is trained using  a batch size of 32, with a  maximum  of 1000 epochs, early stopping with a patience of 10, and a validation split  of 20\%. The model is optimized using the Adam optimizer and the mean squared error loss function. In case of the RF, the number of forest trees is 300, the splitting criterion squared error, the minimum samples split 2 and minimum samples leaf 1. The specific FFNN and RF architectures are determined  through  hyperparameter grid search for the data-driven case. The chosen hyperparameters are maintained consistent  across  all hybrid approaches as well as all documentation/sensor scenario in order to isolate the effects of each experimental condition. \\
In integrating the physics-based and data-driven components, we  explore four approaches: residual, assistant, surrogate and augmentation. For the augmentation approach, fine-tuning is performed  with  early stopping  set to a patience of 3. Fine-tuning of the Augmentation-LR model is achieved by updating the LR weights using the Adam optimizer. The fine-tuning in case of the Augmentation-RF is achieved by using the previously fitted trees on the simulated data as warm start and adding 100 more trees for learning on the real data.\\

\begin{table}[h]
    \centering
    \begin{tabular}{@{}lccc@{}}
    \toprule
    Documentation & \multicolumn{3}{c}{Sensors} \\  \cmidrule{2-4}
          & Weather   & Weather/Building       & Weather/Building/Room     \\ \midrule
    Archetypal              &  \makecell{\textbf{\textit{W-scenario}} \\ Cesar-P} &  -               & -          \\
    Detailed                &  -        &  \makecell{\textbf{\textit{WB-scenario}} \\ Uncalibrated EP}     & \makecell{\textbf{\textit{WBR-scenario}} \\ Calibrated EP} \\
    \bottomrule
    \end{tabular}
    \caption{Documentation and sensor data settings with corresponding physics-based model part and scenario name.}
    \label{table:scenarios}
\end{table}

\begin{table*}[h]
    \centering
    \begin{tabular}{@{}ll@{}}
    \toprule
    Feature group    & Feature variables     \\ \midrule
    Datetime    & Season, week (weekday/weekend), daytime (morning/afternoon/evening/night) \\
    Weather     &   Drybulb \& dewpoint temperature, diffuse \& diffuse solar radiation, rel. humidity, wind direction \& speed\\
    Building    &  Total cooling \& heating mass flows, network temperature, air-conditioning mode (cool/heat) \\
    Room        & Mass flow, temperature setpoint, occupancy, window position (closed/open), blinds position (up/down) \\
    \bottomrule
    \end{tabular}
    \caption{Feature groups with corresponding feature variables. Note that the datetime group is used for all W-, WB- and WBR-scenario.}
    \label{table:features}
\end{table*}

\section{Results and analysis}

\subsection{Prediction performance}
\begin{figure*}[h]
	\centering
	\includegraphics[width=1\linewidth]{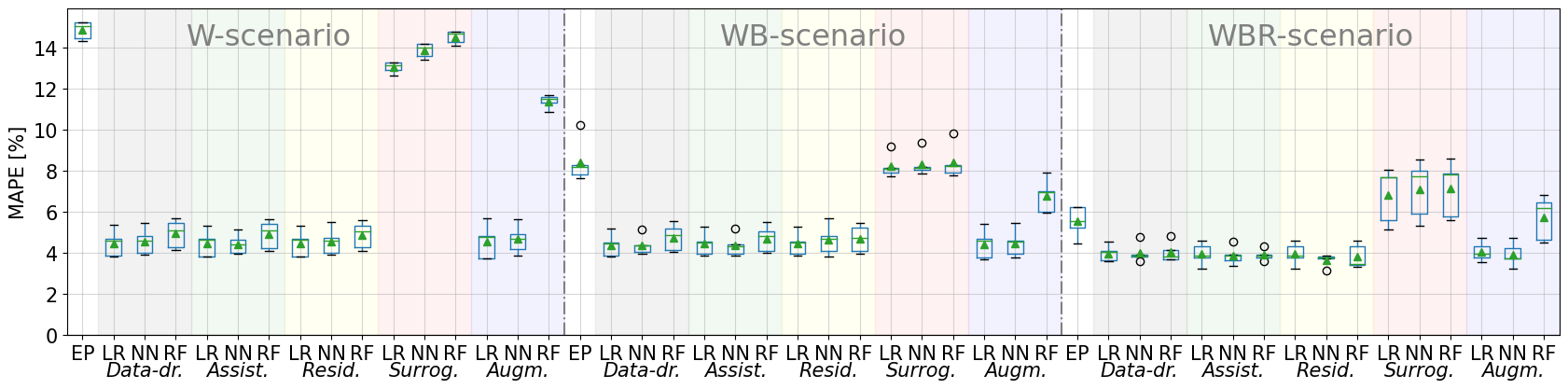}
	\caption{MAPE boxplot for the three scenarios W, WB and WBR between real indoor temperature and model prediction. The green triangle and line denote mean resp. median of the error distribution.}
	\label{fig:wdr_comparison}
\end{figure*}

In order to comprehensively evaluate the hybrid models' prediction performance in real-world settings, we explore their performance in three different documentation/sensor scenarios.
The results for the three documentation/sensor scenarios employing the time series regression framework are illustrated  in Figure~\ref{fig:wdr_comparison}, presented as a MAPE boxplot across all rooms. Across all methods, the average MAPE (indicated by the green triangles) decreases as the level of documentation and sensor data increases from W-scenario to WBR-scenario. Notably, for the best performing method, Residual-FFNN, the MAPE improves by 0.88\%, decreasing from 4.55 \% in the W-scenario to 3.67 \% in the WBR-scenario. This indicates that while the weather features and archetypal building characteristics  provide a robust  foundation  for prediction, significant gains are  achieved through the inclusion of more more sensors and documentation data.
Furthermore, all hybrid methods,except for the surrogate models, outperform  the EP simulator in terms of MAPE. Across all scenarios, the hybrid models exhibit performance comparable to the purely data-driven LR and FFNN. Nevertheless, we can see a slight performance advantage for the Residual-FFNN in the WBR-scenario.

\begin{figure}[h]
	\centering
	\includegraphics[width=1\linewidth]{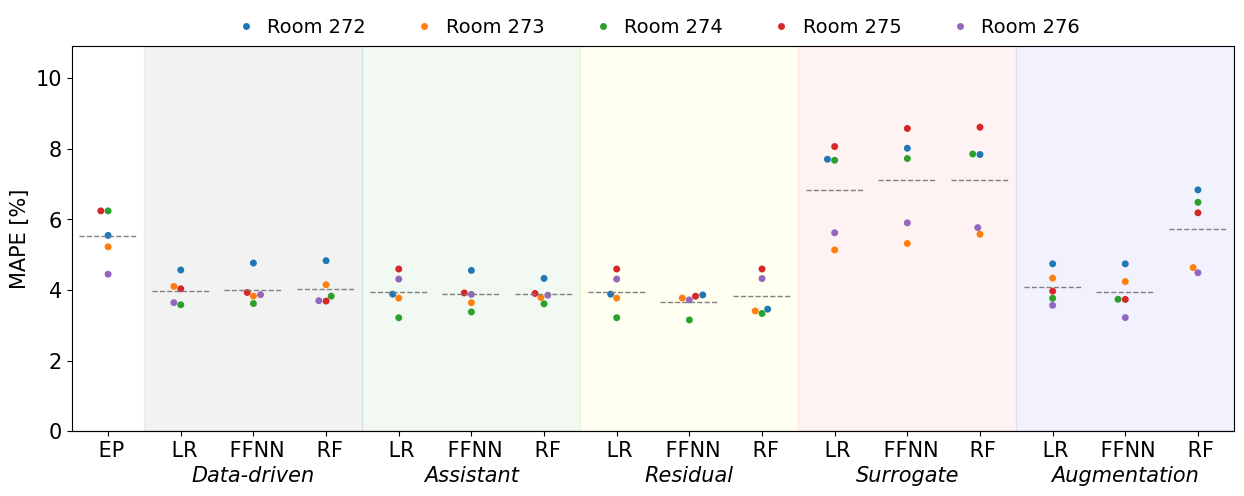}
	\caption{MAPE  for all rooms, hybrid approaches as well as pure physics-based and data-driven models in the WBR-scenario. The grey-dotted line indicates the mean MAPE across all rooms.}
	\label{fig:wdr_scenario}
\end{figure}

\begin{figure}[h]
	\centering
	\includegraphics[width=1\linewidth]{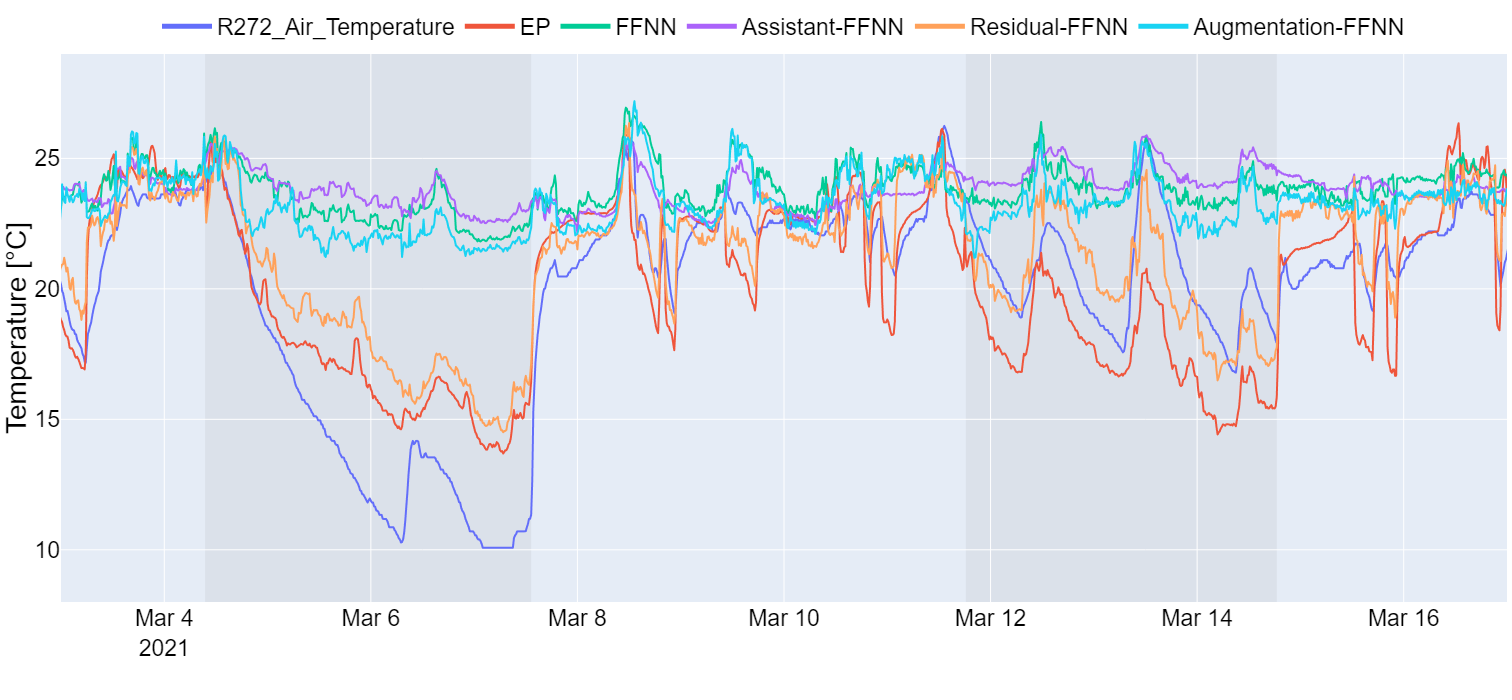}
	\caption{Forecast visualization for a selection of hybrid models for bedroom 272 in the WBR-scenario. The predictions of the surrogate approach are omitted for better visibility. The grey shaded areas indicate the extended period of window openings.}
	\label{fig:why_res}
\end{figure}

Figure~\ref{fig:wdr_scenario}  presents the performance of the WBR scenario, broken down by room. While the Residual-FFNN demonstrates the best overall performance (indicated by the grey-dotted line), there are notable variations  across  different  room types. Specifically, the Residual-FFNN achieves  the lowest MAPE in bedroom 274, whereas the Residual-RF performs better in the living room 273 and bedroom 272. For the two bathrooms, the Augmentation-FFNN yields  the lowest MAPE for bathroom 276 and the data-driven RF the best MAPE performance for bathroom 275. 
An example of the Residual-FFNN's superior performance is highlighted  in Figure~\ref{fig:why_res}. During extended window openings from March  4th to 7th and March 11th to 15th  (shaded in grey),  the Residual-FFNN accurately captures the extreme temperature fluctuations by leveraging the EP model.  In contrast, the purely data-driven model and other hybrid approaches fail to predict this behaviour effectively. The FFNN shows difficulties to learn relationships that generalize to such extreme window behaviour. The Assistant-FFNN approach seems unable to beneficially use the EP simulation as feature, even deteriorating performance compared to FFNN in these cases. In the augmentation approach, we observe a slight improvement over FFNN, indicating that the two-step learning approach can bring advantages in such situations. A similar effect can be observed in the residual approaches of LR and RF Figure~\ref{fig:why_res_LR} and Figure ~\ref{fig:why_res_RF}. The mean performance of all methods for MAPE, MAE and RMSE is given in Figure~\ref{tab:metrics}.

\begin{figure}[h]
	\centering
	\includegraphics[width=1\linewidth]{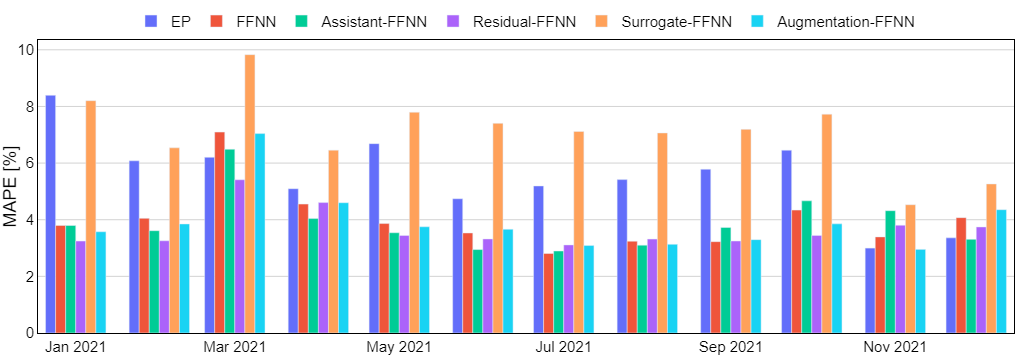}
	\caption{Average MAPE across rooms and grouped by month for the hybrid FFNN models.}
	\label{fig:monthly_error}
\end{figure}

Figure~\ref{fig:monthly_error} presents  the average MAPE across rooms, grouped by month, for the hybrid models. Two notable observations can be made. 
Firstly, March and April exhibit  a higher error rate, exceeding  4 \% MAPE for all models. This is due to extended periods of open windows in the bedrooms and living room, which cause a drop in room temperature and trigger the heating system to compensate. These extreme fluctuations  are challenging for all models to accurately capture.
Secondly, in November and December, the EP model performs  competitively with the other models  compared to the rest of the year. An explanation for this is the rental situation of UMAR, as the unit was uninhabited in winter 2021. Consequently, the temperature patterns in November and December 2021 were more  regular, making it easier  for the EP model to simulate. In contrast, the other models were  trained on data from  2020 when the unit was inhabited during  winter. This makes it harder for them to generalize to an uninhabited period.

\subsection{Influence of historical sensor data}
\begin{figure}[h]
	\centering
	\includegraphics[width=1\linewidth]{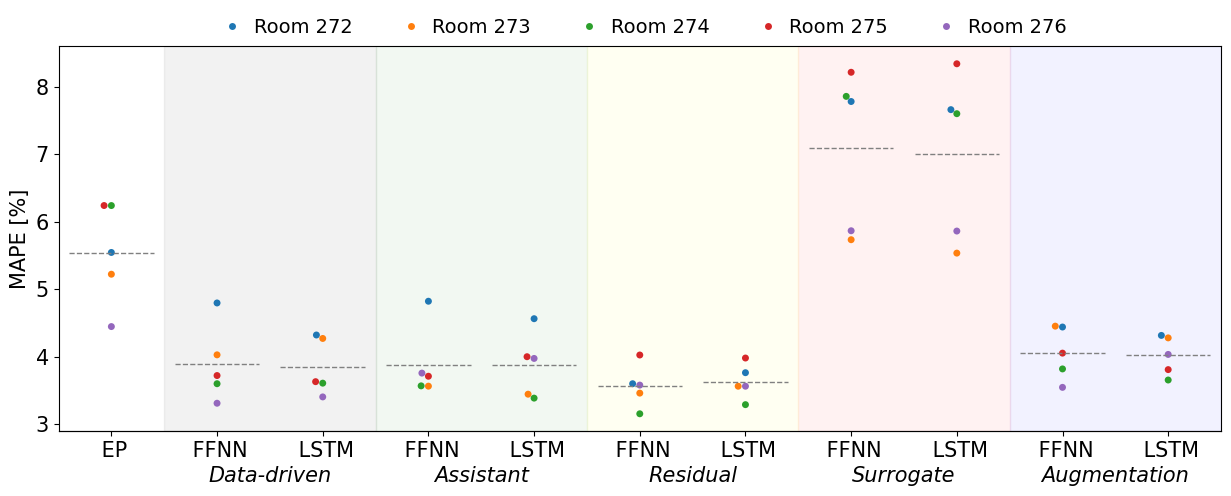}
	\caption{Comparison between hybrid approaches of FFNN and LSTM for the case of one hour history integration.}
	\label{fig:hour_history}
\end{figure}

The influence of historical sensor data can give insights into the impact of short- and long-term effects on prediction performance. To this end, we examine a Long Short-term Memory (LSTM) Network \cite{hochreiter_long_1997}. To be comparable to the FFNN, we employ a two layer network with 128 neurons each and train it in the same fashion.\\
In Figure \ref{fig:hour_history}, we display the MAPE for the hybrid approaches with FFNN and LSTM. We observe that the Residual approach shows the best overall performance, with a slight advantage of Residual-FFNN over Residual-LSTM. This could indicate that the residual approach has a superior ability to leverage short-term historical effects. Nevertheless, when looking at the room-wise performance, we see that the pure data-driven models outperform the residual approach for the two bathrooms. This phenomenon could be traced back again to the peculiar dynamics of the bathrooms, complicating their physics-based modeling.

\begin{figure}[h]
	\centering
	\includegraphics[width=1\linewidth]{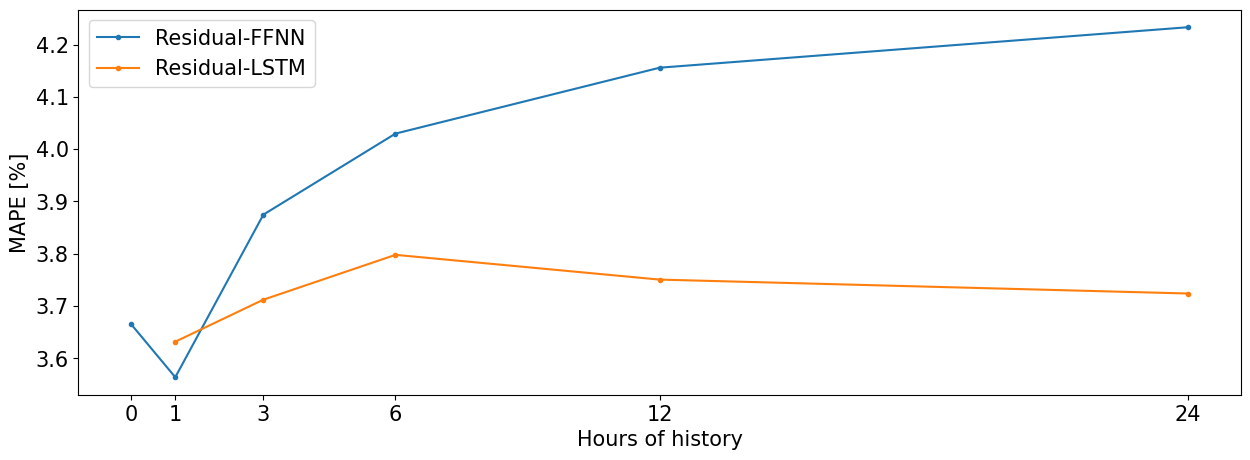}
	\caption{Comparison between Residual-FFNN and Residual-LSTM when including different hours of sensor history. Note that 0 hours of history represents the direct input-output mapping from features to target at the same time step.}
	\label{fig:residual_history}
\end{figure}

Figure \ref{fig:residual_history} shows the MAPE performance of the Residual-FFNN and Residual-LSTM for increasing hours of history. In order to isolate the effect of history integration, the model architectures are kept constant. We observe that the lowest error is achieved by the Residual-FFNN with 1 hour of historical sensor data. Interestingly, the Residual-LSTM also displays its best performance with 1 hour history integration. This could indicate that short-term effects in UMAR such as solar radiation are predominant, and integrating longer-term effects may not bring additional benefits. Further, we observe a large performance degradation with increasing hours for Residual-FFNN. This suggests an underfitting of the model, since one additional hour of history corresponds to an addition of about 200 features. Conversely, the Residual-LSTM has a superior ability to integrate the historical data having a substantially lower error for increasing hours.

\subsection{Explainability within model}

\begin{figure}[h]
	\centering
	\includegraphics[width=0.8\linewidth]{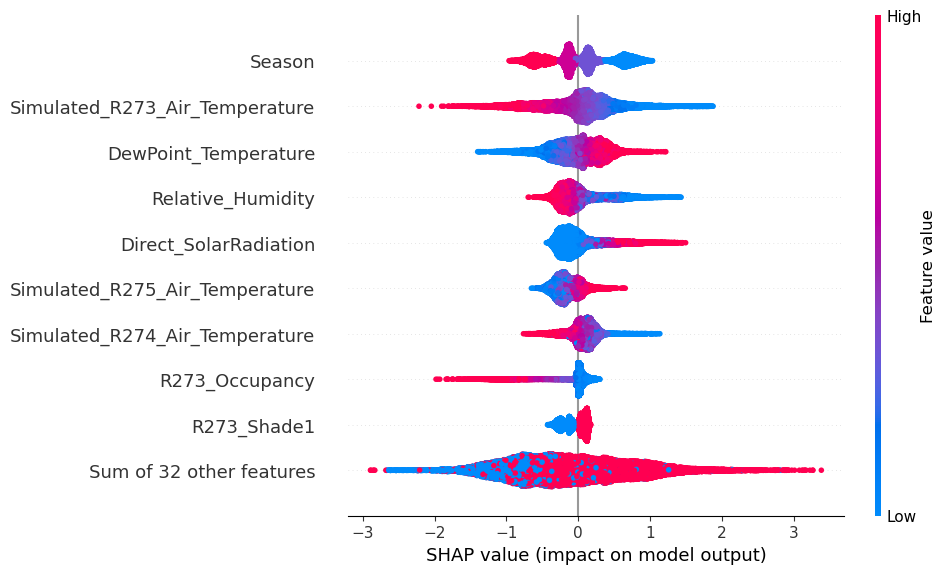}
	\caption{SHAP beeswarm plot with 10 most influential features for living room 273 for the Residual-FFNN in the WBR-scenario.}
	\label{fig:beeswarm}
\end{figure}

In order to improve the understanding of the data-driven model as well as of the interdependencies between the physics-based and data-driven model, we investigate the models by using hierarchical Shapley Value framework to explain their predictions. 
Our analysis focuses on the WBR-scenario due the highest documentation and sensor availability level. Moreover, we concentrate on living room 273 because it's the largest room and on the Residual-FFNN given its best average performance across all rooms. Figure~\ref{fig:beeswarm} presents  the SHAP beeswarm plot for the hierarchical Shapley values, highlighting  the 10 most influential features for living room 273 as identified by the Residual-FFNN model. Among the top features are the simulated EP input, weather variables (solar radiation, relative humidity, dew-point temperature), as well as room-specific variables  like occupancy and shading. It is intuitive that the simulated EP inputs play a key role in  a residual approach. Additionally, the significant contribution of room-level variables underscores their importance in model performance. The  high importance of weather variables suggests that the residual approach compensates  for  weather variations not fully captured by the EP model.

\begin{figure}[h]
	\centering
	\includegraphics[width=0.8\linewidth]{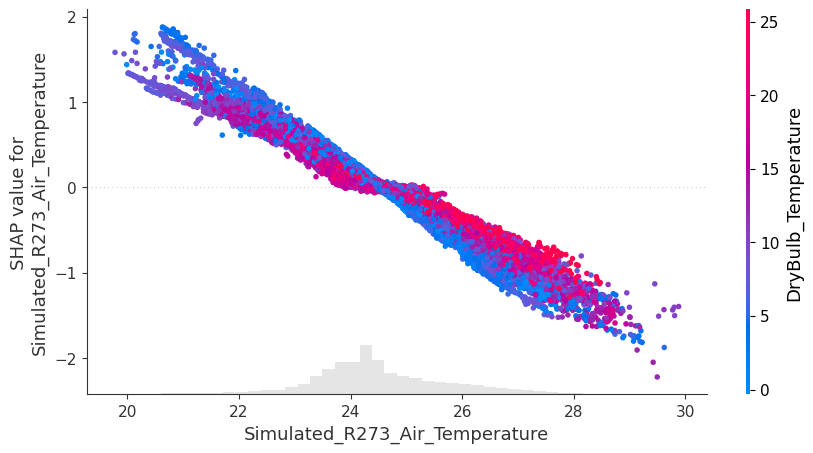}
	\caption{SHAP dependence plot for living room 273 according to Residual-FFNN showing the dependence between the simulated EP input and the corresponding Shapley value colored by dry-bulb temperature.}
	\label{fig:dependence}
\end{figure}

Figure~\ref{fig:dependence} presents the SHAP dependence plot for living room 273, highlighting  the second most important feature (Simulated air temperature) according to Figure~\ref{fig:beeswarm}. Note that the Shapley value represents the average contribution to the output and in this case the contribution in temperature degrees to the residual correction. The plot reveals that when  the simulated temperature  input exceeds a threshold of  24.6\textdegree C, the contribution to the residual correction is generally negative. Below this threshold, the contribution to the correction is positive. Examining  the dry-bulb temperature coloring, we observe  that  contributions to residual corrections for higher temperatures tend to be negative and up to a correction contribution of -2 degrees, which may suggest a bias in the EP model at  higher temperatures. This insight could serve as a basis for improving the EP model.

\begin{figure}[h]
	\centering
	\includegraphics[width=0.8\linewidth]{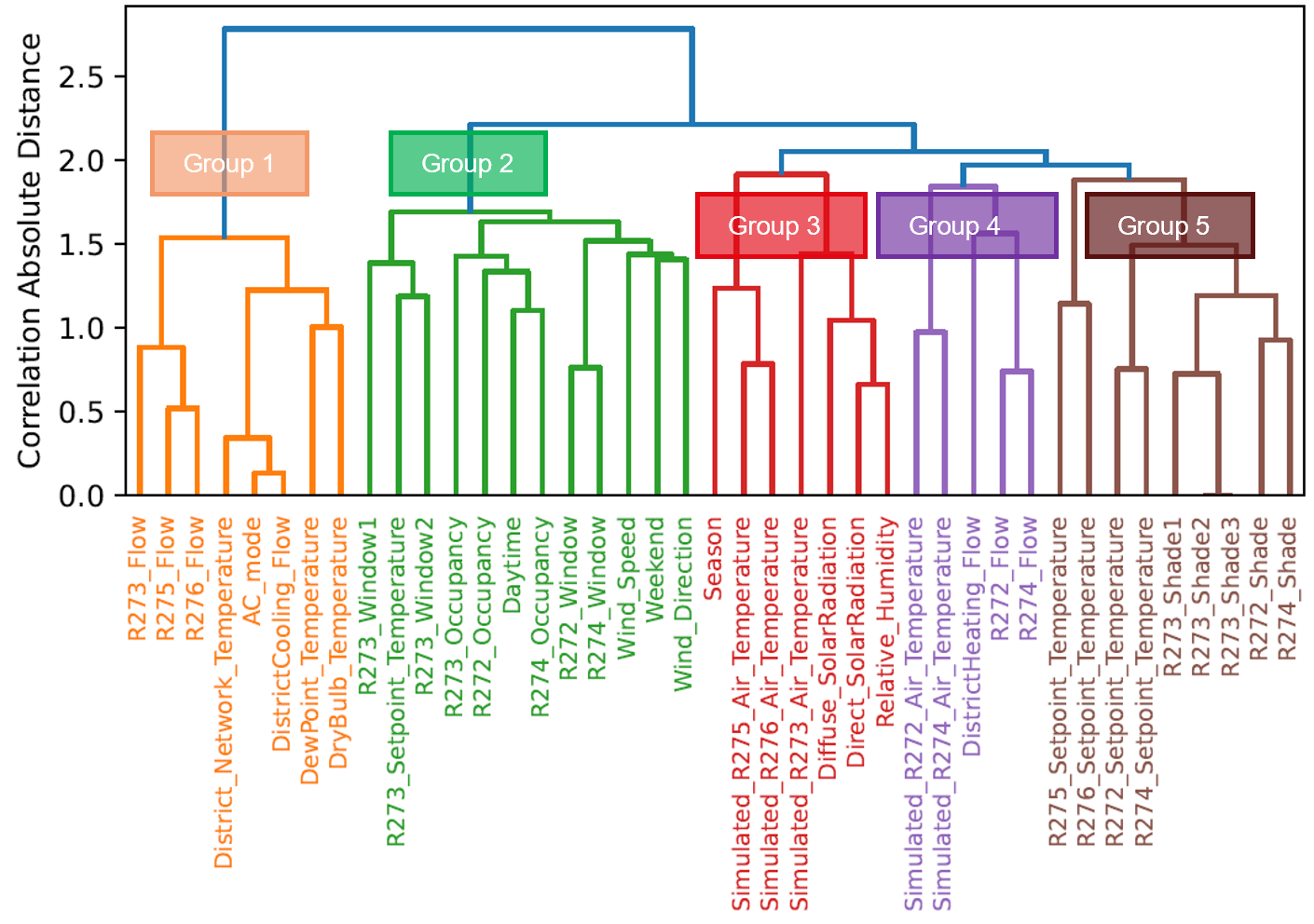}
	\caption{Dendrogram for input of Residual-FFNN in the WBR-scenario.}
	\label{fig:dendogram_ar_group}
\end{figure}

Figure~\ref{fig:dendogram_ar_group} displays the hierarchical Shapley dendrogram for the inputs of the best-performing method, Residual-FFNN. With a few minor exceptions, four distinct groups can be identified. The orange cluster is composed mainly of outdoor temperature and room mass flow variables. The green cluster includes  wind, window, and occupancy variables. The red and violet clusters are dominated by  EP simulated inputs, solar radiation, and relative humidity, while the brown group consists of setpoints and shading variables. Notably, Table ~\ref{tab:hierarchical_mean} exposes that the red cluster holds the highest relative importance among all clusters for room 273. This indicates that solar radiation, simulated air temperatures and relative humidity are the most impactful input group. The strong influence of solar radiation is consistent with the observation that the living room has three large windows. At the same time, the importance of the simulated air temperatures can be attributed to the nature of the residual model. If we cut the dendrogram at the second level, the red, violet and brown groups together have an aggregated hierarchical Shapley value  that is more than double that of the combined orange and green clusters for room 273. We can interpret features directly related to indoor temperature (simulated air temperatures and temperature set points) as well as irradiation related features (solar radiation and shading) contribute significantly more on average than outside temperature, wind features, window openings and occupancy. A similar effect is observed for the two bedrooms, except that the violet group has gained importance due to the simulated temperature of the respective bedrooms being in this group. In case of the bathrooms, we notice a decrease of importance in all groups besides the red cluster, of which the simulated temperatures of bathrooms are part of. Such an effect can be attributed to the more isolated dynamics of the bathrooms having no windows.

\begin{table}[h]
\centering
\begin{tabular}{@{}lrrrrr@{}}
\toprule
                & \cellcolor{orange!35}Group 1 & \cellcolor{green!35}Group 2 & \cellcolor{red!35}Group 3 & \cellcolor{violet!35}Group 4 & \multicolumn{1}{l}{\cellcolor{brown!35}Group 5} \\ \midrule
Bedroom 272     & \cellcolor{orange!35}0.61                        & \cellcolor{green!35}0.40                        & \cellcolor{red!35}0.84                        & \cellcolor{violet!35}0.77                        & \multicolumn{1}{r}{\cellcolor{brown!35}0.75}                        \\
Living room 273 & \cellcolor{orange!35}0.73                        & \cellcolor{green!35}0.49                       & \cellcolor{red!35}1.55                        & \cellcolor{violet!35}0.36                        & \multicolumn{1}{r}{\cellcolor{brown!35}0.75}                        \\
Bedroom 274     & \cellcolor{orange!35}0.47                        & \cellcolor{green!35}0.38                       & \cellcolor{red!35}0.96                        & \cellcolor{violet!35}0.80                         & \multicolumn{1}{r}{\cellcolor{brown!35}0.52}                        \\
Bathroom 275    & \cellcolor{orange!35}0.20                         & \cellcolor{green!35}0.27                       & \cellcolor{red!35}0.88                        & \cellcolor{violet!35}0.13                        & \multicolumn{1}{r}{\cellcolor{brown!35}0.33}                        \\
Bathroom 276    & \cellcolor{orange!35}0.30                         & \cellcolor{green!35}0.20                        & \cellcolor{red!35}0.69                        & \cellcolor{violet!35}0.17                        & \multicolumn{1}{r}{\cellcolor{brown!35}0.36}                        \\ \midrule
Room average    & \cellcolor{orange!35}0.46                         & \cellcolor{green!35}0.35                        & \cellcolor{red!35}0.98                        & \cellcolor{violet!35}0.45                        & \multicolumn{1}{r}{\cellcolor{brown!35}0.54}                       \\\bottomrule
\end{tabular}
\caption{Hierarchical Shapley values for all dendrogram groups in the case of the Residual-FFNN.}
    \label{tab:hierarchical_mean}
\end{table}

\subsection{Explainability comparison between models}
\begin{figure}[h]
	\centering
	\includegraphics[width=1.0\linewidth]{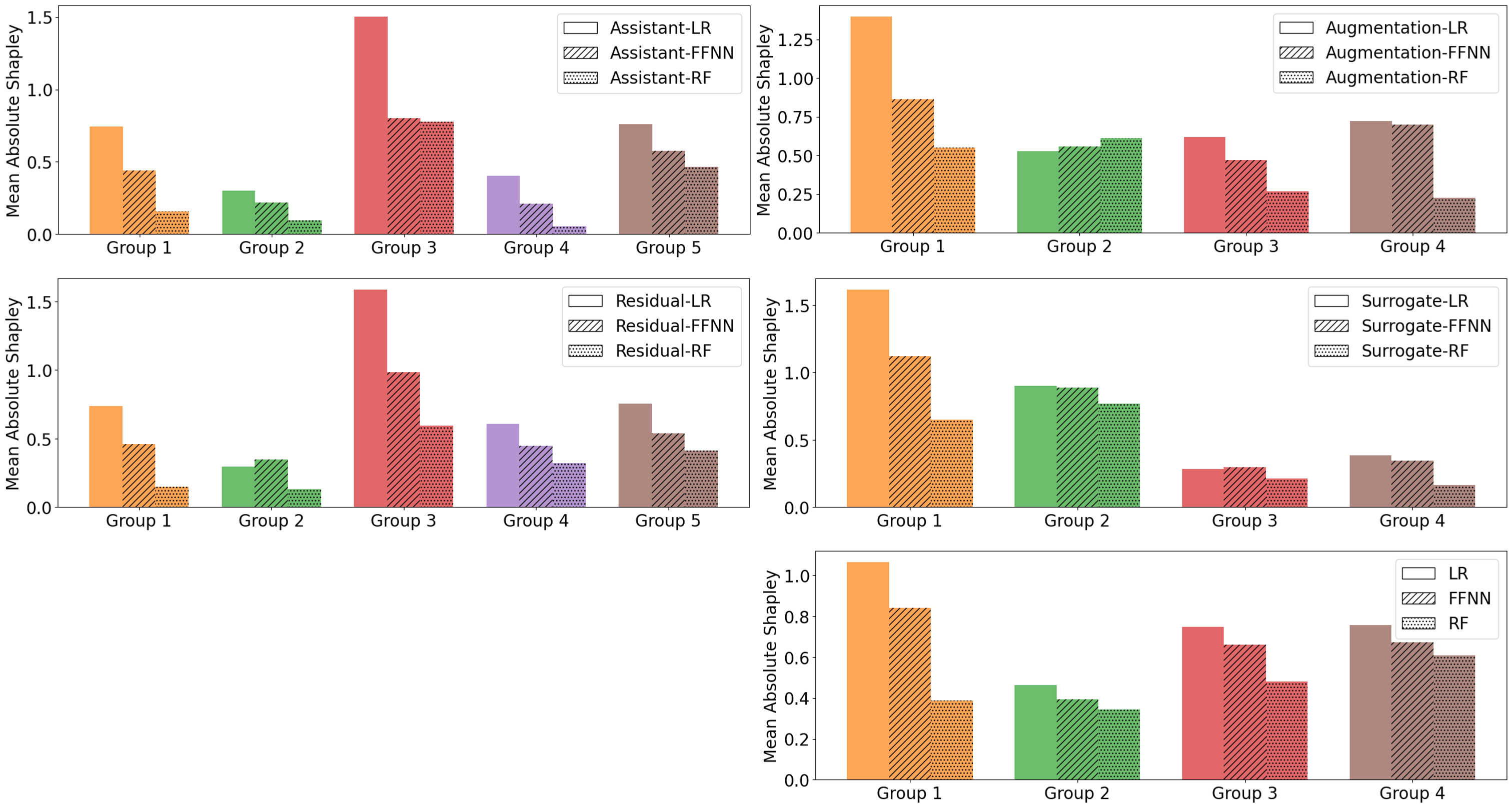}
	\caption{Averaged absolute hierarchical Shapley value across rooms for each cluster group and hybrid approach. Note that values between different models should only be compared relative to each other.}
	\label{fig:group_shap_comparison}
\end{figure}

In order to compare the influences of the input cluster groups across the different hybrid approaches, Figure ~\ref{fig:group_shap_comparison} shows the averaged absolute hierarchical Shapley values across rooms for LR, FFNN and RF. Note that the cluster groups of the assistant and residual approaches follow the same form shown in Figure~\ref{fig:dendogram_ar_group} due to the additional simulated input, and are henceforth denoted as Assisted/Residual-clustering (AR-clustering). The data-driven, augmentation and surrogate approach follow the same form given in Figure~\ref{fig:dendogram_cas_group} having no additional simulated input and are denoted as Data-driven/Augmentation/Surrogate-clustering (DAS-clustering). In principle, the DAS-clustering resembles the AR-clustering for most parts, except for the absence of simulated inputs, the violet group being dissolved into the orange and green one as well as two set point input being merged into the green cluster. Moreover, note that the values between LR, FFNN, and RF hybrid approaches should not be compared directly but only in relative terms, since the Shapley value calculation is highly dependent on the model's inner workings.
The most informative observation in Figure ~\ref{fig:group_shap_comparison} is the high importance of the red cluster relative to the other clusters in case of the assistant and residual approaches for RL, FFNN and RF. In contrast, the DAS-clustering bar plots show a more balanced picture with the orange group having the highest importance. This suggests that the assistant and residual approach are provided with most parts of the dynamics through the simulated inputs in the red group, while the data-driven, augmentation and surrogate approach have to learn the thermodynamics primarily through the sensor data.

It is also instructive to compare the native importance measures of LR, FFNN and RF with the hierarchical Shapley values. In case of LR, the linear effects are computed as the regression coefficients multiplied by the features for each sample. Linear effects are very simple to calculate, but do not account for feature interactions. A way to quantify importance for the FFNN represents the Jacobian sensitivity, where the change of the output with respect to the input is calculated for each sample. The Jacobian reflects the direct influence of each input to output, but does not capture feature interactions and assumes a local linear approximation of the input-output relationship. For the RF, we can calculate the mean decrease in impurity calculated as the total reduction of the splitting criterion induced by a specific feature. A drawback of the mean decrease impurity calculation is that features with many unique values can be overly favored since they have more opportunities to reduce the impurity. Moreover, the mean decrease impurity does not account for feature interactions.
A ranking comparison of the five most important features for the model native importance measures and the hierarchical Shapley values in the case of the residual approach is given in Table~\ref{tab:model_native_vs_shap}. We see that in case of the linear effects of LR and Jacobian sensitivity of FFNN, three respectively four out of five features actually overlap in the top 5 ranking. In contrast, the mean impurity decrease of RF has a perfect overlap for the top 5 positions. We can conclude that even though the model native importance measures do not account for feature interactions, the top features are relatively well captured. Furthermore, the mean impurity decrease of RF seems to be the most reliable measure in this regard. The complete feature importance plots can be found in Figure~\ref{fig:shap_res_lr_vs_effects}, Figure~\ref{fig:shap_res_ffnn_vs_sensitivity} and Figure~\ref{fig:shap_res_rf_vs_impurity}.

\begin{table}[h]
\centering
\begin{tabular}{llrr}
\hline
\multicolumn{1}{r}{}                                                                   &        & Model Native                      & Hierarchical Shapley              \\ \hline
\multirow{5}{*}{\begin{tabular}[c]{@{}l@{}}Linear Effect -\\ LR\end{tabular}}          & Rank 1 & Simulated\_R275\_Air\_Temperature & Simulated\_R275\_Air\_Temperature \\
& Rank 2 & Simulated\_R276\_Air\_Temperature & Simulated\_R276\_Air\_Temperature \\
& Rank 3 & Relative\_Humidity                & DewPoint\_Temperature             \\
& Rank 4 & R274\_Shade                       & Simulated\_R274\_Air\_Temperature \\
& Rank 5 & DewPoint\_Temperature             & Simulated\_R272\_Air\_Temperature \\ \hline
\multirow{5}{*}{\begin{tabular}[c]{@{}l@{}}Jacobian Sensitivity -\\ FFNN\end{tabular}} & Rank 1 & Simulated\_R272\_Air\_Temperature & Simulated\_R275\_Air\_Temperature \\
& Rank 2 & Simulated\_R276\_Air\_Temperature & Season                            \\
& Rank 3 & Simulated\_R273\_Air\_Temperature & Simulated\_R272\_Air\_Temperature \\
& Rank 4 & Simulated\_R275\_Air\_Temperature & Simulated\_R274\_Air\_Temperature \\
& Rank 5 & Season                            & Simulated\_R273\_Air\_Temperature \\ \hline
\multirow{5}{*}{\begin{tabular}[c]{@{}l@{}}Gini Impurity -\\ RF\end{tabular}}          & Rank 1 & Simulated\_R276\_Air\_Temperature & Season                            \\
& Rank 2 & Simulated\_R272\_Air\_Temperature & Simulated\_R276\_Air\_Temperature \\
& Rank 3 & Season                            & Simulated\_R272\_Air\_Temperature \\
& Rank 4 & DistrictHeating\_Flow             & Simulated\_R275\_Air\_Temperature \\
& Rank 5 & Simulated\_R275\_Air\_Temperature & DistrictHeating\_Flow             \\ \hline
\end{tabular}
\caption{Model native importance measures contrasted with hierarchical Shapley values for Residual-LR, -FFNN and -RF}
\label{tab:model_native_vs_shap}
\end{table}

\subsection{Impact of data quantity on performance}
\begin{figure}[h]
	\centering
	\includegraphics[width=1\linewidth]{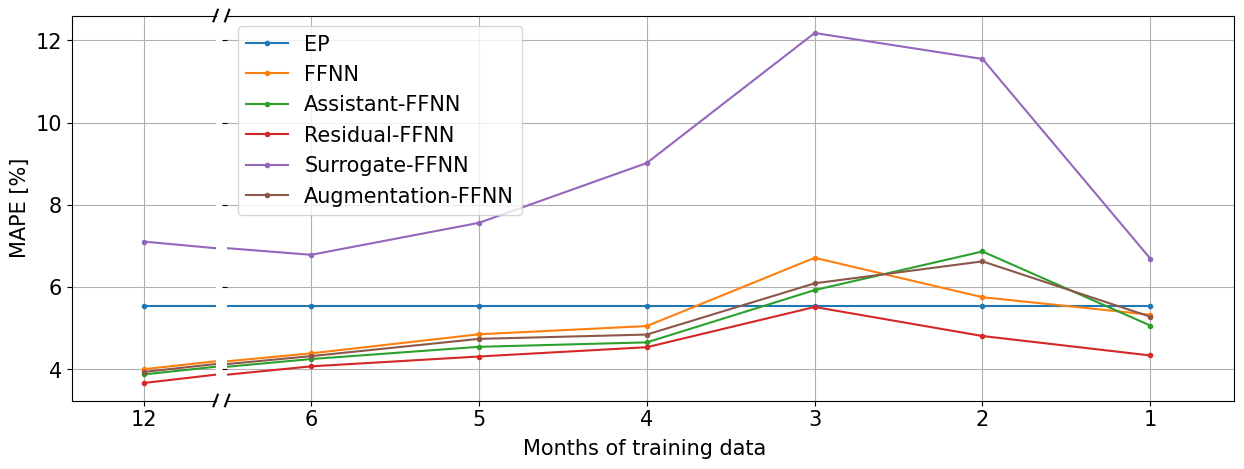}
	\caption{Average MAPE across all rooms for number of months of training data for all methods in the WBR-scenario.}
	\label{fig:data_quantity}
\end{figure}

\begin{figure}[h]
	\centering
	\includegraphics[width=1\linewidth]{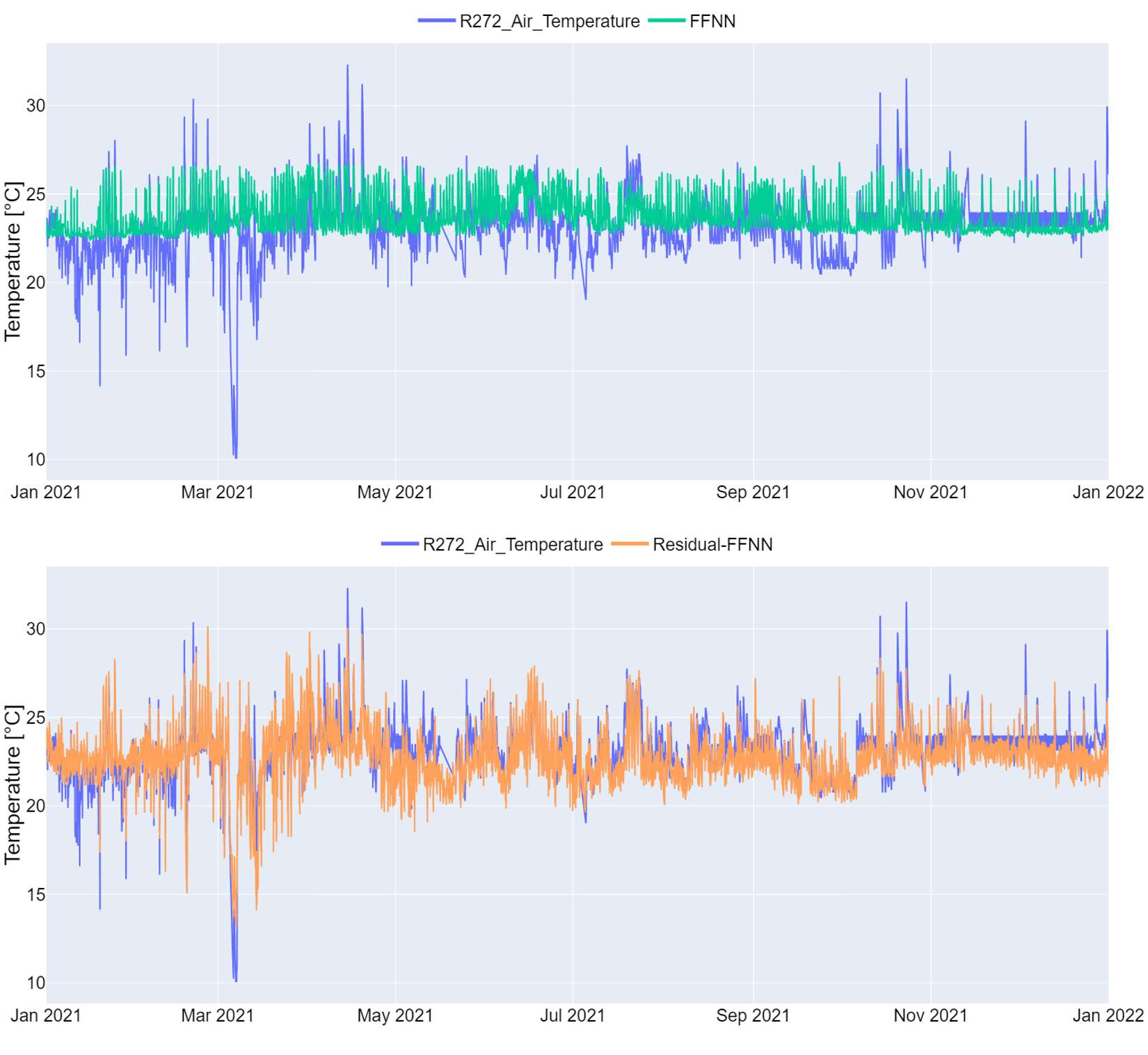}
	\caption{Predictions of FFNN and Residual-FFNN over the test set in the case of 1 month of training data.}
	\label{fig:1m_ffnn_predictions}
\end{figure}

In practice, building sensor data may not be abundantly available due to recent sensor installation, temporary sensor deployment or sensor failures. In this light, we investigate the data quantity dependence of the hybrid models with limited data. In these experiments, we reduce the amount of training data and evaluate the hybrid models on the same test set (year 2021) as before. The MAPE results for these experiments, using limited training data, are displayed in Figure~\ref{fig:data_quantity}. The models are trained with a progressively decreasing number of months, ranging from January 15th to  July 15th. The results for 12 months of training data  are shown as a reference on the left side of the figure. For all experiments, we keep the same calibration of the EnergyPlus model as in the 12-months set-up, assuming ideal conditions.\\
Overall, we observe a decline in performance as  the amount of training data decreases, with two notable exceptions at 3 months and 1 month. The 3-months performance is heavily  influenced by the validation set,  which predominantly consists of data collected in April. Since April has the highest indoor mean temperature and standard deviation, it does not serve as a  representative validation set. This results in more biased models and poorer performance on the test set. The 1-month performance,  however, is an evaluation artifact. Although the predictions cover less variance in the true room temperature, they are closer to the true temperature mean, resulting in a slightly better MAPE than when training with 2 months of data.\\
Among the models, the Residual-FFNN performs best, while the Surrogate-FFNN consistently shows the highest MAPE across  all  training  data sizes. Additionally,  the performance gap between the FFNN and Residual-FFNN widens as the amount of training data increases. Notably, from three months of training data, the limited data significantly impacts the FFNN's performance. In contrast, the Residual-FFNN effectively  leverages the physical information from the EP simulation to generate more accurate predictions. An illustration of this can be seen in Figure~\ref{fig:1m_ffnn_predictions}, which shows predictions over the entire test set using only 1 month of training data. Similar effects can be observed in cases of LR and RF, depicted in Appendix Figure~\ref{fig:data_quantity_lr}, Figure~\ref{fig:data_quantity_rf}, Figure~\ref{fig:1m_lr_predictions} and Figure~\ref{fig:1m_rf_predictions}. In summary, we see that the performance declines with decreasing training data while the residual approach maintains the best ability to cover the temperature variations.

\section{Discussion} 
Our results indicate  that more detailed  documentation for the physics-based model and increased sensor availability significantly improve  prediction accuracy.  In scenarios with limited documentation and sensor data, hybrid approaches do not show substantial  improvements over purely data-driven models.  However, in scenarios with comprehensive documentation and abundant sensor data, hybrid methods demonstrate considerable performance enhancements. This suggests that hybrid approaches reach their full  potential only when  sufficient sensor measurements related to thermodynamics are available.
Additionally, we observe  that the most influential inputs vary depending on the  hybrid approach used. For hybrid methods incorporating physics-based simulations as additional inputs (assistant and residual approaches), the feature group containing simulation data exhibits  the highest relevance, indicating  that the physics-based simulations provide the majority of information about thermodynamics. In contrast, for hybrid approaches like augmentation and surrogate, the importance is more evenly distributed across feature groups, indicating that these methods learn the dynamics directly from sensor data. \\
In scenarios with the highest documentation and sensor availability, different hybrid models  perform best for different room types. However, the Residual-FFNN consistently achieves the best average performance across all rooms compared to the other hybrid models. We also find that the residual approach  is most effective at capturing out of distribution behaviour, such as extended periods of window openings. Even in scenarios with limited data, the residual approach successfully  captures temperature variations, whereas purely data-driven model and other hybrid approaches struggle. We attribute  these advantages to the robust  foundation provided by the physics-based model, combined with data-driven corrections. In stark contrast, the surrogate approach  consistently  performs the worst across all experiments, likely due to the physics-based model's performance setting  an upper bound. \\
A limiting factor in the analysis of the limitations of training data is the assumption of ideal calibration of the physics-based model. With limited data  available for calibration, the process may be suboptimal, leading to degraded performance of the physics-based model. We hypothesize that a residual-based approach could compensate for such systematic errors, whereas  other hybrid approaches, such as surrogate models, may experience more significant performance limitations. Further studies are required to investigate the effects of calibration under these constrained conditions.\\
When applying the hierarchical Shapley value with the Residual-FFNN, we showcase the potential to detect biases in the physics-based component and identify  areas for correction. Specifically, the feature importances in the residual approach show  increased  contributions to  residual corrections at high outside temperatures, indicating  that the physics-based model exhibits  significant  bias under these conditions.\\
For real-world deployment in control applications, the primary challenges are expected to include handling  sensor faults, managing missing data, and adapting to usage behaviour changes due to shifting  occupancy patterns. Compared to a purely physics-based model, the hybrid model is anticipated to have a minimal  computational burden at inference. In selecting a suitable hybrid model, various criteria must be considered. The residual approach is best suited for compensating systematic biases in the physics-based model, such as deviations resulting from calibration errors or weather variability. The assisted approach similarly incorporates the physics-based simulation as feature, but its reliance on data-driven training means that the contribution of the physics-based simulation is determined by the model itself, offering practitioners less control. The surrogate model, in contrast, is primarily constrained by the performance of the underlying physics-based model making it less suitable for similar operation settings. The augmentation approach refines the surrogate model through fine-tuning on real data, potentially improving accuracy. However, finding an optimal tuning strategy can be complex and resource-intensive.

\section{Conclusion}
In this study, we assess the performance of four predominant hybrid approaches along with purely data-driven and physics-based models for predicting building temperatures. To mimic challenging real-world conditions, we evaluate these hybrid methods across three common scenarios that vary in documentation detail and sensor availability. Furthermore, we explore the explainability of the hybrid approaches with the hierarchical Shapley value to account for feature dependencies. The results show that the residual approach is the most effective hybrid approach for temperature prediction, while the surrogate approach consistently exhibits the lowest accuracy. In terms of explainability, we showcased the adequacy of the hierarchical Shapley value to interpret the hybrid approaches.
While the residual approach proved to be most effective in our settings, further improvements could involve weighting samples based on a certainty metric for correction. This would enable the residual model to prioritize time periods where higher error corrections are more significant.
Future development of our study could involve conducting experiments on a larger scale, including  a greater number of buildings and various  building types beyond residential. In addition, while  we considered multiple  data scenarios, further investigation into data availability could involve more granular levels of building documentation.
Another area  for improvement is the assumption of ideal physics-based calibration when assessing  the impact of limited training data on performance. We plan to address this by re-calibrating the physics-based model based on available data. An alternative research direction could explore an online setting for hybrid approaches, incorporating frequent re-calibration of the physics-based model as well as continuous re-training of the data-driven model.

\printcredits

\section*{Data availability}
The code and dataset are publicly available at the \href{https://github.com/Leo-VK/hybrid_bem}{GitHub repository}.

\section*{Acknowledgement}
We thank all involved members from the Urban Energy Systems Laboratory at Empa and Intelligent Maintenance and Operations Systems Laboratory at EPFL. Special thanks go to Fazel Khayatian for insightful discussions about Building Energy Modeling and the nestli platform. This project is funded by Empa research \& development grant 5213.00276.

\bibliographystyle{elsarticle-num}

\bibliography{cas-refs}

\clearpage
\appendix

\section{Additional results}

\FloatBarrier
\begin{table}[htbp]
\centering
\begin{tabular}{@{}lrrr@{}}
\toprule
Method            & MAPE [\%]     & MAE [\textdegree C]   & RMSE [\textdegree C]    \\ \midrule
EP                & 5.54          & 1.25          & 1.59          \\
LR                & 3.98          & 0.92          & 1.26          \\
FFNN              & 4.00          & 0.92          & 1.26          \\
RF                & 4.04          & 0.93          & 1.31          \\
Assistant-LR      & 3.95          & 0.91          & 1.21          \\
Assistant-FFNN    & 3.87          & 0.89          & 1.23          \\
Assistant-RF      & 3.90          & 0.90          & 1.27          \\
Residual-LR       & 3.95          & 0.91          & 1.21          \\
Residual-FFNN     & \textbf{3.67} & \textbf{0.85} & \textbf{1.14} \\
Residual-RF       & 3.82          & 0.89          & 1.21          \\
Surrogate-LR      & 6.84          & 1.53          & 1.89          \\
Surrogate-FFNN    & 7.11          & 1.59          & 1.94          \\
Surrogate-RF      & 7.13          & 1.59          & 1.95          \\
Augmentation-LR       & 4.07          & 0.94          & 1.27          \\
Augmentation-FFNN     & 3.93          & 0.91          & 1.25          \\ 
Augmentation-RF       & 5.72          & 1.28          & 1.62          \\\bottomrule
\end{tabular}
\caption{Average MAPE, MAE and RMSE across all rooms for each method in the WBR-scenario}
\label{tab:metrics}
\end{table}

\begin{figure}[h]
	\centering
	\includegraphics[width=1\linewidth]{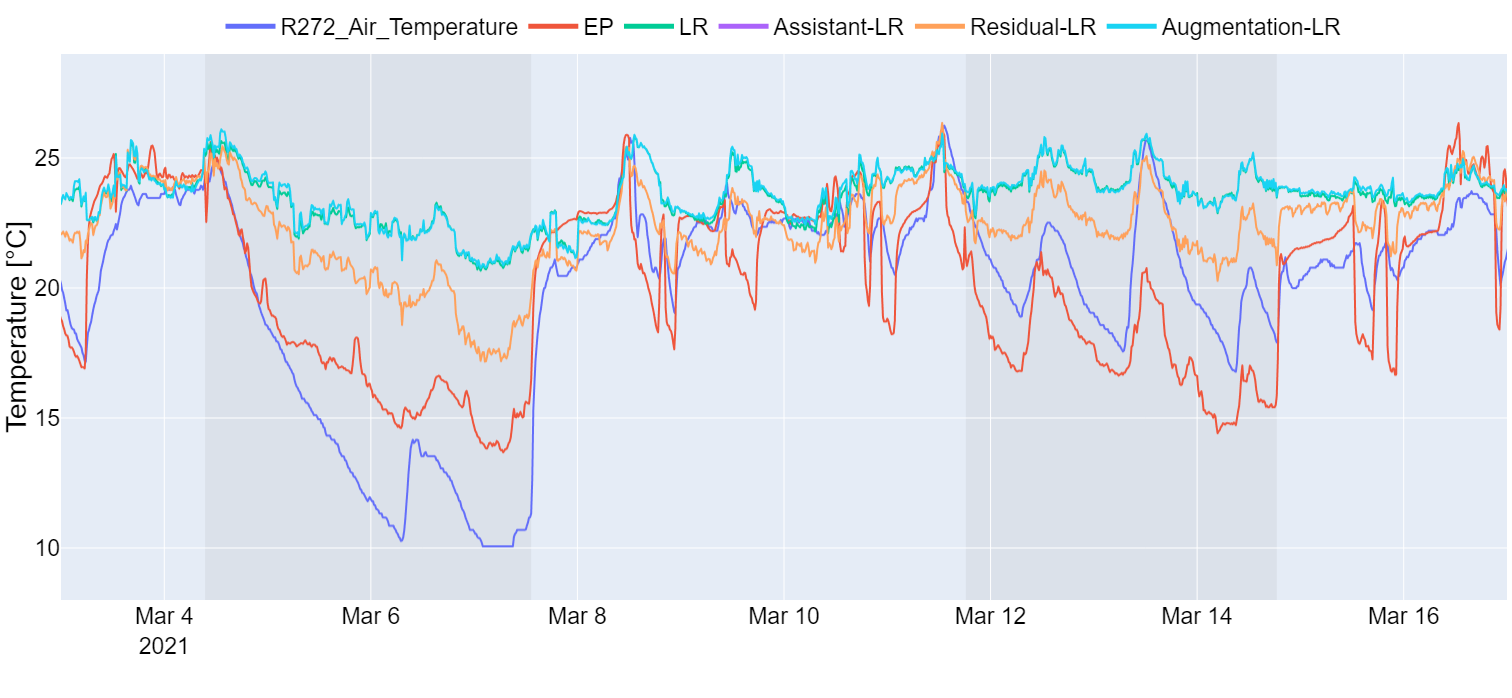}
	\caption{Forecast visualization for a selection of hybrid models in the case of LR for bedroom 272 in the WBR-scenario. The predictions of the surrogate approach are omitted for better visibility. The grey shaded areas indicate the extended period of window openings.}
	\label{fig:why_res_LR}
\end{figure}

\begin{figure}[h]
	\centering
	\includegraphics[width=1\linewidth]{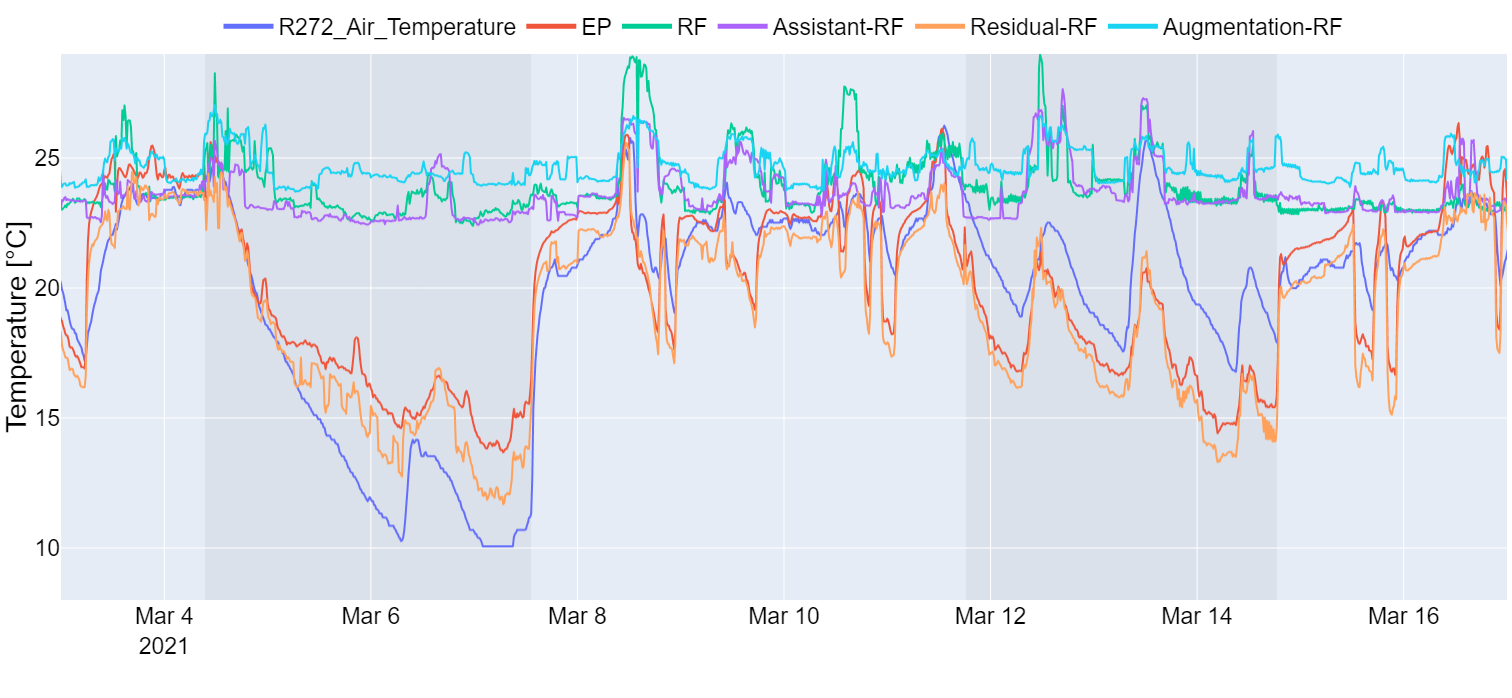}
	\caption{Forecast visualization for a selection of hybrid models in the case of RF for bedroom 272 in the WBR-scenario. The predictions of the surrogate approach are omitted for better visibility. The grey shaded areas indicate the extended period of window openings.}
	\label{fig:why_res_RF}
\end{figure}

\begin{figure}[h]
	\centering
	\includegraphics[width=1\linewidth]{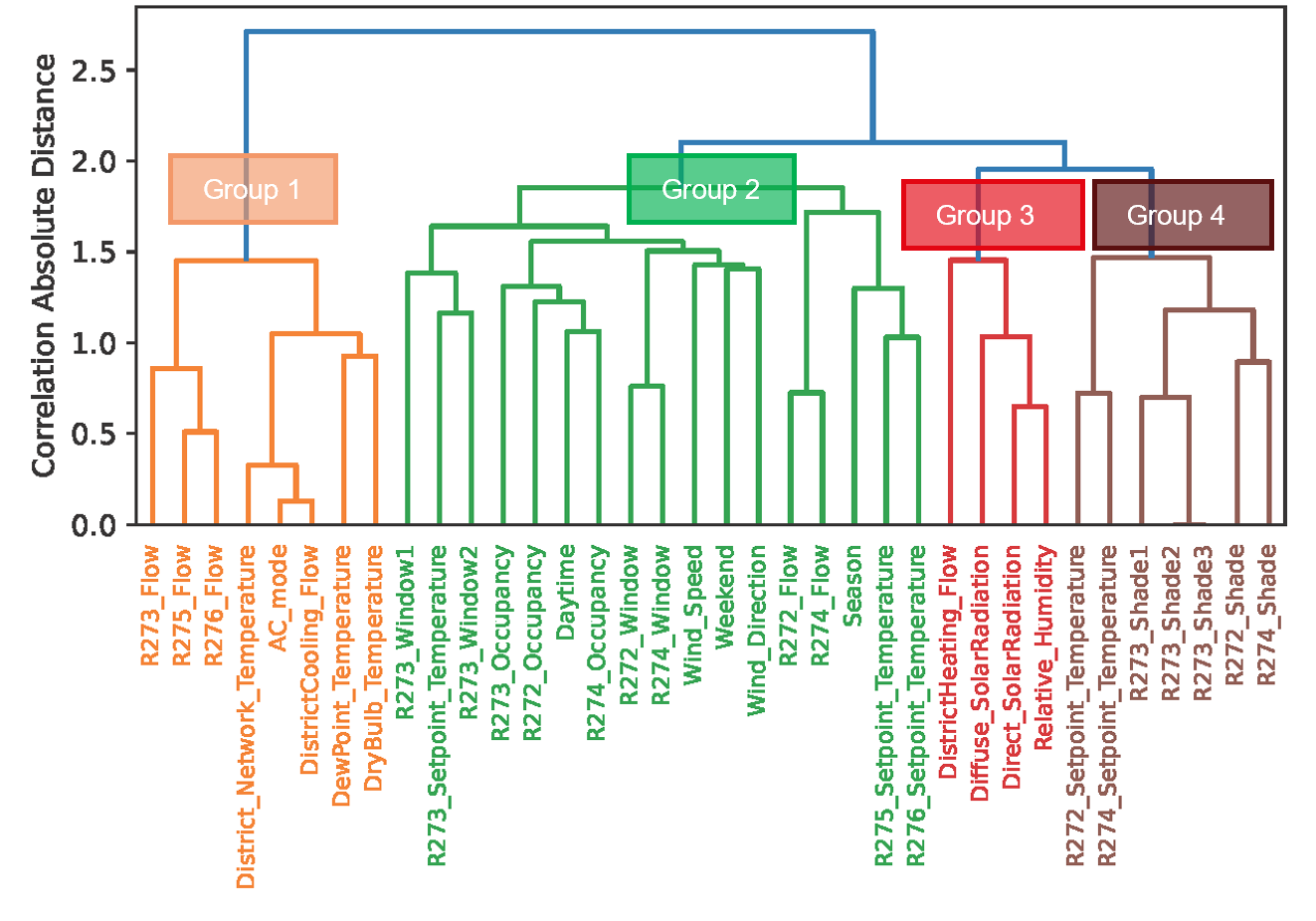}
	\caption{Dendrogram for input of data-driven, augmentation, and surrogate approach in the WBR-scenario.}
	\label{fig:dendogram_cas_group}
\end{figure}

\begin{figure}[h]
	\centering
	\includegraphics[width=1\linewidth]{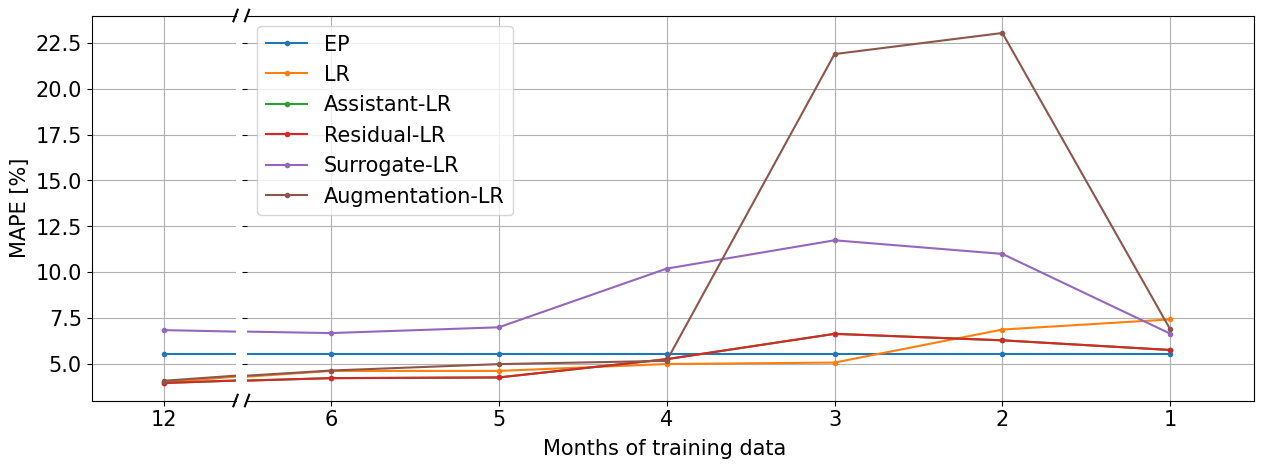}
	\caption{Average MAPE for LR across all rooms for number of months of training data for all methods in the WBR-scenario.}
	\label{fig:data_quantity_lr}
\end{figure}

\begin{figure}[h]
	\centering
	\includegraphics[width=1\linewidth]{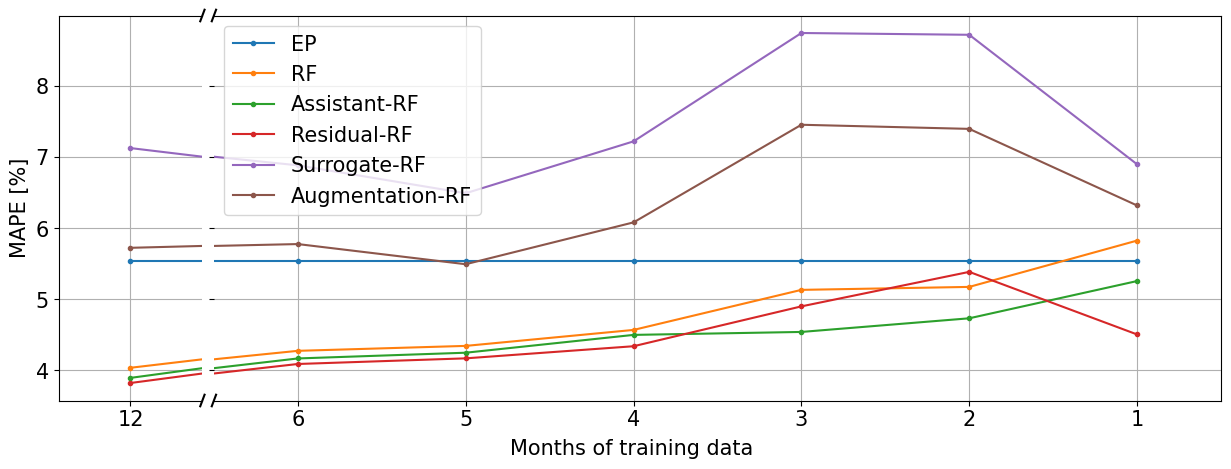}
	\caption{Average MAPE for RF across all rooms for number of months of training data for all methods in the WBR-scenario.}
	\label{fig:data_quantity_rf}
\end{figure}

\begin{figure}[h]
	\centering
	\includegraphics[width=1\linewidth]{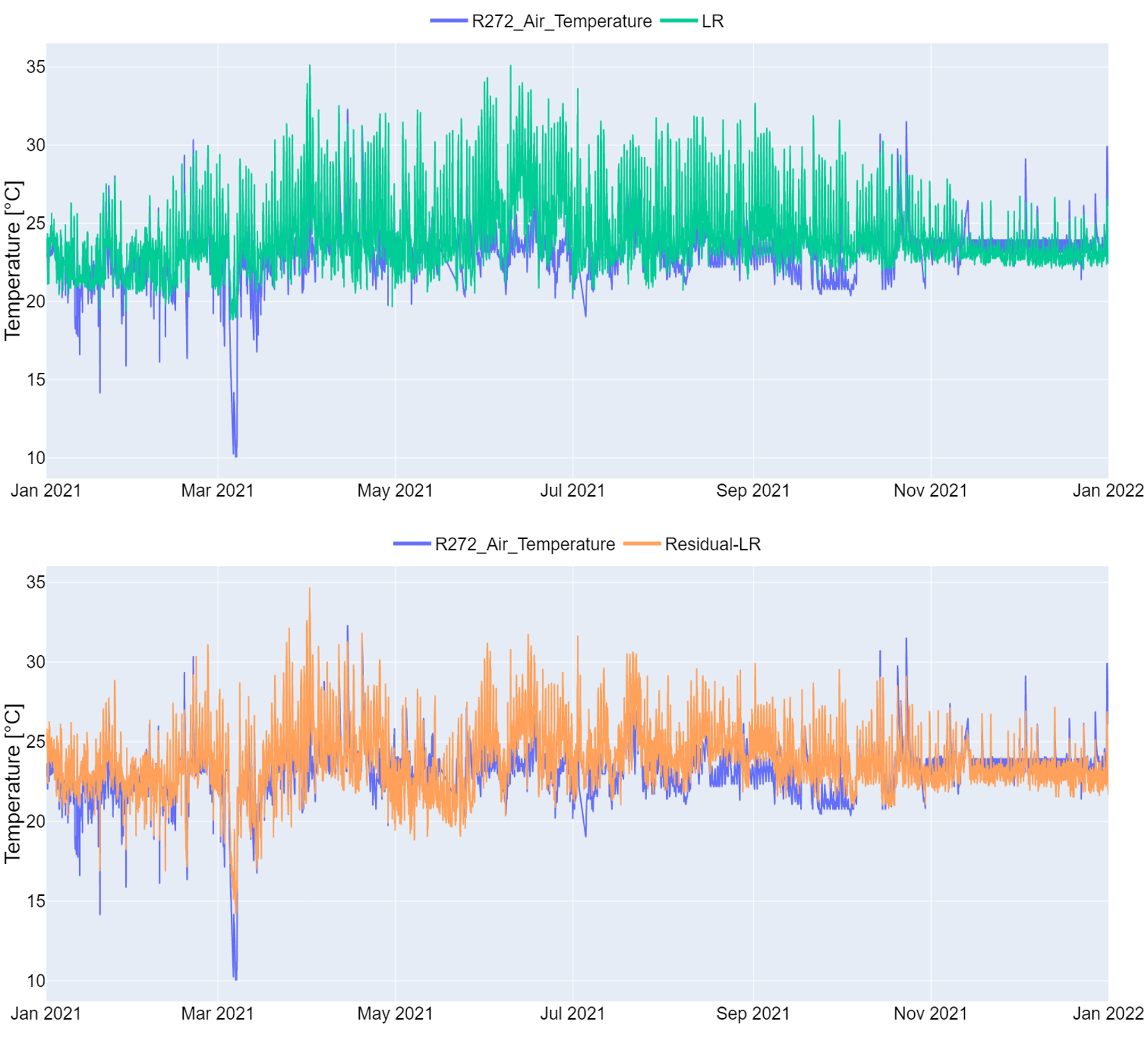}
	\caption{Predictions of LR and Residual-LR over the test set in the case of 1 month of training data.}
	\label{fig:1m_lr_predictions}
\end{figure}

\begin{figure}[h]
	\centering
	\includegraphics[width=1\linewidth]{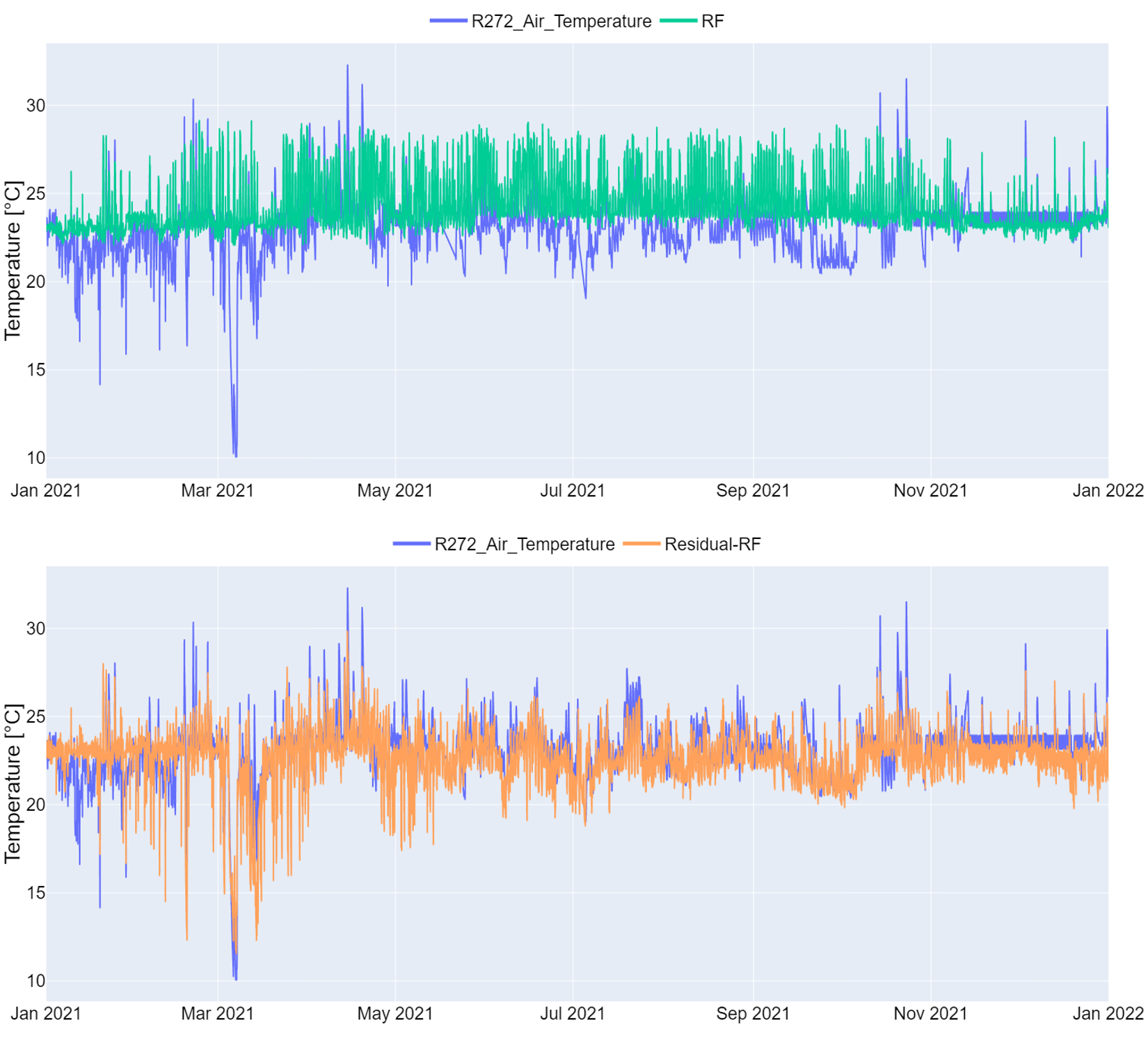}
	\caption{Predictions of RF and Residual-RF over the test set in the case of 1 month of training data.}
	\label{fig:1m_rf_predictions}
\end{figure}

\begin{figure}[h]
	\centering
	\includegraphics[width=1\linewidth]{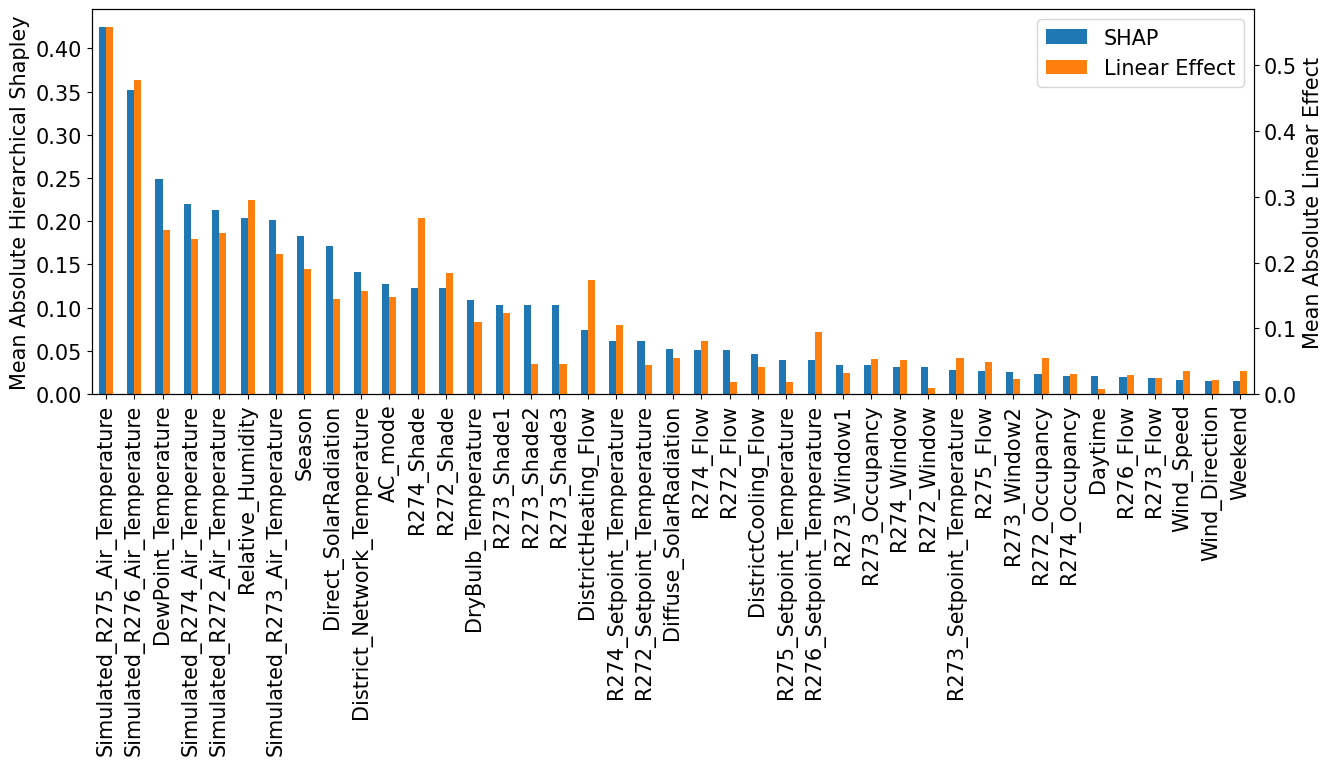}
	\caption{Bar plot of the mean absolute linear effects and hierarchical Shapley values for all features in the case of Residual-LR.}
	\label{fig:shap_res_lr_vs_effects}
\end{figure}

\begin{figure}[h]
	\centering
	\includegraphics[width=1\linewidth]{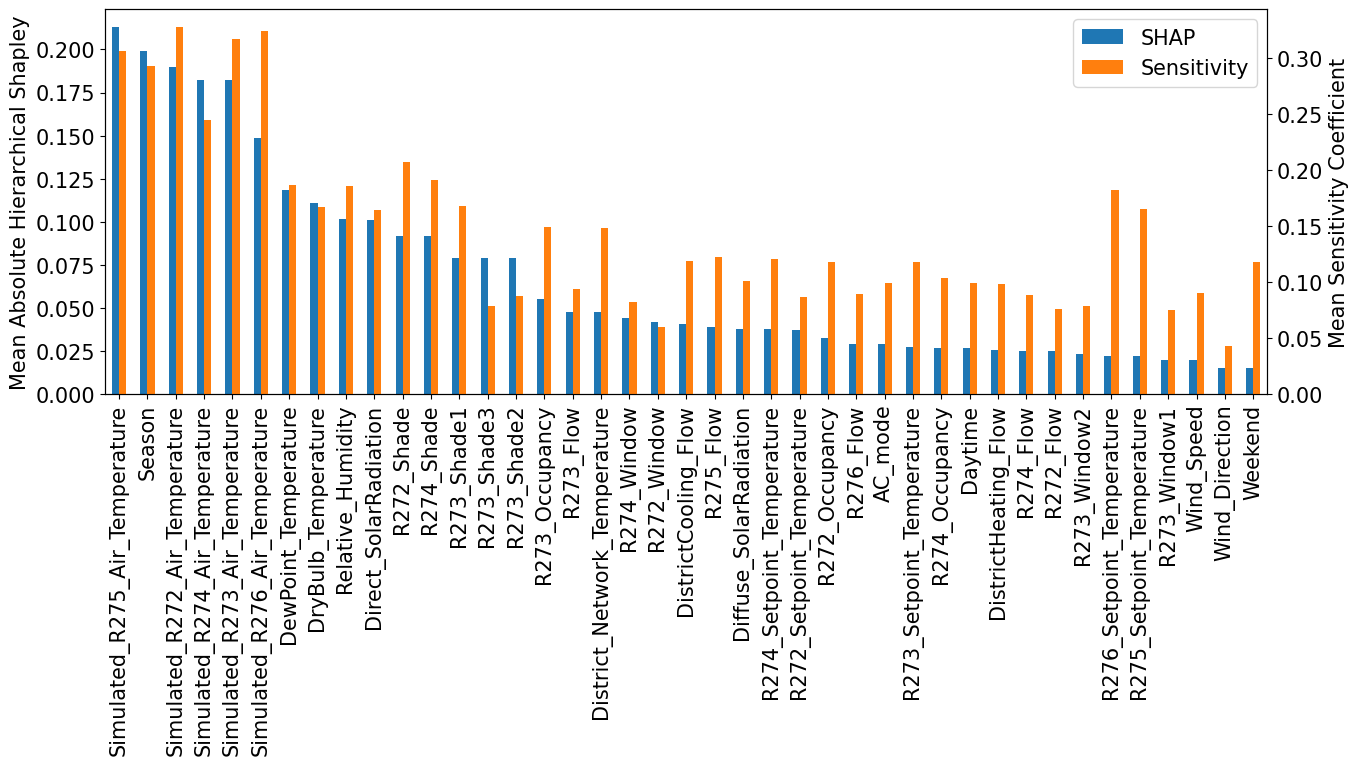}
	\caption{Bar plot of the mean absolute sensitivity coefficients and hierarchical Shapley values for all features in the case of Residual-FFNN.}
	\label{fig:shap_res_ffnn_vs_sensitivity}
\end{figure}

\begin{figure}[h]
	\centering
	\includegraphics[width=1\linewidth]{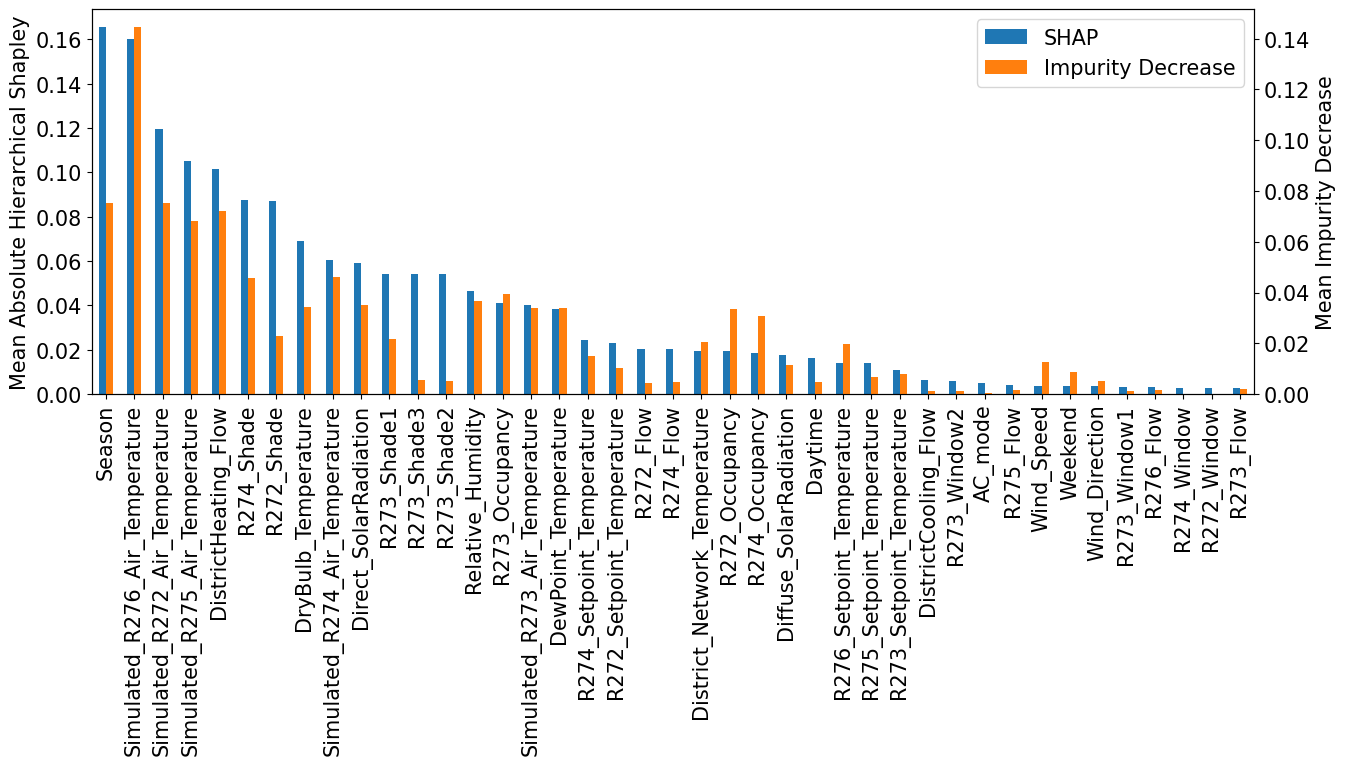}
	\caption{Bar plot of the mean impurity decrease and hierarchical Shapley values for all features in the case of Residual-RF.}
	\label{fig:shap_res_rf_vs_impurity}
\end{figure}

\end{document}